\documentclass[a4paper]{article}
\usepackage{amsmath,amssymb,amsfonts,graphicx,cite}
\def\bequ{\begin{equation}}
\def\eequ{\end{equation}}
\def\half{{1\over 2}}

\def\ben{\begin{equation}}
\def\een{\end{equation}}
\def\half{{\textstyle{1\over2}}}

\def\nn{\nonumber} \def\bd{\begin{document}} \def\ed{\end{document}}
\def\ds{\documentstyle} \let\fr=\frac \let\bl=\bigl \let\br=\bigr
\let\Br=\Bigr \let\Bl=\Bigl
\let\bm=\bibitem
\let\na=\nabla
\let\pa=\partial \let\ov=\overline
\def\be{\begin{equation}}
\def\ee{\end{equation}}
\def\ba{\begin{array}}
\def\ea{\end{array}}
\def\ft#1#2{{\textstyle{\frac{\scriptstyle #1}{\scriptstyle #2} } }}
\def\fft#1#2{{\frac{#1}{#2}}}
\def\del{\partial}
\def\vp{\varphi}
\def\sst#1{{\scriptscriptstyle #1}}
\def\oneone{\rlap 1\mkern4mu{\rm l}}
\def\td{\tilde}
\def\wtd{\widetilde}
\def\ie{{\it i.e.\ }}
\def\dalemb#1#2{{\vbox{\hrule height .#2pt
        \hbox{\vrule width.#2pt height#1pt \kern#1pt
                \vrule width.#2pt}
        \hrule height.#2pt}}}
\def\square{\mathord{\dalemb{6.8}{7}\hbox{\hskip1pt}}}
\newcommand{\ho}[1]{$\, ^{#1}$}
\newcommand{\hoch}[1]{$\, ^{#1}$}
\newcommand{\bea}{\begin{eqnarray}}
\newcommand{\eea}{\end{eqnarray}}
\newcommand{\ra}{\rightarrow}
\newcommand{\lra}{\longrightarrow}
\newcommand{\Lra}{\Leftrightarrow}
\newcommand{\bp}{\tilde \beta^\prime}
\newcommand{\Tr}{{\rm Tr} }
\def\0{{\sst{(0)}}}
\def\1{{\sst{(1)}}}
\def\2{{\sst{(2)}}}
\def\3{{\sst{(3)}}}
\def\4{{\sst{(4)}}}
\def\5{{\sst{(5)}}}
\def\6{{\sst{(6)}}}
\def\7{{\sst{(7)}}}
\def\8{{\sst{(8)}}}
\def\n{{\sst{(n)}}}
\def\cA{{{\cal A}}}
\def\cB{{{\cal B}}}
\def\cF{{{\cal F}}}
\def\cG{{{\cal G}}}
\def\cH{{{\cal H}}}
\def\cL{{{\cal L}}}

\def\tV{\widetilde V}
\def\tW{\widetilde W}
\def\tH{\widetilde H}
\def\tE{\widetilde E}
\def\tF{\widetilde F}
\def\tA{\widetilde A}
\def\im{{{\rm i}}}
\def\tY{{{\wtd Y}}}
\def\ep{{\epsilon}}
\def\vep{{\varepsilon}}
\def\bD{{{\bar D}}}
\def\R{{{\mathbb R}}}
\def\C{{{\mathbb C}}}
\def\H{{{\mathbb H}}}
\def\CP{{{\mathbb C}{\mathbb P}}}
\def\RP{{{\mathbb R}{\mathbb P}}}
\def\Z{{{\mathbb Z}}}
\def\bA{{{\mathbb A}}}
\def\bB{{{\mathbb B}}}
\def\bC{{{\mathbb C}}}
\def\bD{{{\mathbb D}}}
\def\bE{{{\mathbb E}}}
\def\bZ{{{\mathbb Z}}}
\def\Re{{{\frak{Re}}}}
\def\Im{{{\frak{Im}}}}
\def\cosec{{\,\hbox{cosec}\,}}
\def\Gm{{\Gamma_{\!\! -}}}
\def\Gp{{\Gamma_{\!\! +}}}
\def\stan{{standard }}
\def\nonstan{{supernumerary }}
\def\p{{\partial}}
\def\kdel#1{{\fft{\del}{\del#1}}}
\def\Dslash{\slash \negthinspace \negthinspace \negthinspace \negthinspace D}

\def\bog{{Bogomolny }}
\def\om{{\omega}}

\newcommand\w[1]{\\[0.#1cm]}
\def\eq#1{(\ref{#1})}
\def\c{{\gamma}}

\input amssym.def
\input amssym.tex
\def\ben{\begin{equation}}
\def\een{\end{equation}}
\def\half{{1 \over 2}}
\def\bea{\begin{eqnarray}}
\def\eea{\end{eqnarray}}
\def\p{\partial}
\def\m{\mu}
\def\n{\nu}
\def\S{\Sigma}
\def\a{\alpha}
\def\b{\beta}
\def\c{\gamma}
\def\d{\delta}
\def\l{\lambda}
\def\br{{\bf r}}
\def\bt{{\bf t}}
\def\bn{{\bf n}}
\def\bb{{\bf b}}
\def\pa{\partial}
\def\bx{{\bf x}}
\def\ba{{\bf a}}
\def\bv{{\bf v}}
\def\bom{{\mbox{\boldmath $ \omega $ }}} 
\def\bF{{\bf F}}
\def\bG{{\bf G}}
\def\bb{{\bf b}}
\def\by{{\bf y}}
\def\bt{{\bf t}}
\def\bu{{\bf u}}
\def\bp{{\bf p}}
\def\bL{{\bf L}}
\def\bB{{\bf B}}
\def\bE{{\bf E}}
\def\bH{{\bf H}}
\def\bD{{\bf D}}
\def\bS{{\bf S}}
\def\bk{{\bf k}}
\def\bc{{\bf c}}

\begin{document}

\hfuzz=100pt
\title{Measures on Mixing Angles}
\author{Gary W. Gibbons$^1$, Steffen Gielen$^1$, C. N. Pope$^{1,2}$ and Neil Turok$^{1,3}$
\\
\\ {\it $^1$D.A.M.T.P., Centre for Mathematical Sciences, Cambridge University,}
\\ {\it Wilberforce Road, Cambridge, CB3 0WA, U.K.}
\\
\\ {\it $^2$George P. \& Cynthia W. Mitchell Institute for Fundamental Physics and Astronomy,}
\\ {\it Texas A\&M University, College Station, TX 77843-4242, USA}
\\
\\ {\it $^3$Perimeter Institute for Theoretical Physics,} 
\\ {\it 31 Caroline St. N., Waterloo, Ontario, Canada N2L 2Y5}
}

\maketitle
                                                   
\begin{abstract} 

We address the problem of the apparently very small magnitude of CP violation in the standard model, measured by the Jarlskog invariant $J$. In order to make statements about probabilities for certain values of $J$, we seek to find a natural measure on the space of Kobayashi-Maskawa matrices, the double quotient $U(1)^2 \backslash SU(3) / U(1)^2$. We review several possible, geometrically motivated choices of the measure, and compute expectation values for powers of $J$ for these measures. We find that different choices of the measure generically make the observed magnitude of CP violation appear finely tuned. Since the quark masses and the mixing angles are determined by the same set of Yukawa couplings, we then do a second calculation in which we take the known quark mass hierarchy into account. We construct the simplest measure on the space of $3\times 3$ Hermitian matrices which reproduces this known hierarchy. Calculating expectation values for powers of $J$ in this second approach, we find that values of $J$ close to the observed value are now rather likely, and there does not seem to be any fine-tuning. Our results suggest that the choice of Kobayashi-Maskawa angles is closely linked to the observed mass hierarchy. We close by discussing the corresponding case of neutrinos.
\\
\\PACS numbers: 12.15.Hh, 14.60.Pq, 02.20.Hj, 02.40.-k

\end{abstract}

\newpage

\tableofcontents

\newpage

\section{Introduction}

A traditional attitude to theoretical physics has been that the
main problem is to discover the fundamental laws of physics and 
leave it to experiment and observation to decide  what
particular implementation best describes ``Our Universe.''
Thus traditionally a ``physical theory''  is often thought of in terms
of a  
local Lagrangian including   certain ``coupling constants,''
``mass ratios,'' and ``mixing angles,'' all of   which, 
since  Planck's introduction of Planck units \cite{Planck}, may be taken to be 
dimensionless numbers. In addition, the local Lagrangian
must be supplemented with an account of the general class of boundary
conditions for which the variational principle is valid.
Different classes of boundary conditions are usually thought of as
different ``superselection sectors''  of the theory,
and describe qualitatively different types of situations  
which traditionally  are not thought of as having any relation 
to one another.  

Within each sector, there are many solutions of the equations of motion,
each of which may be specified by providing suitable ``initial conditions.''
Classically these conditions may be thought of as the space of 
classical histories,
and given in terms of Cauchy data modulo the   relation
that two sets of Cauchy data giving the same history are taken to be 
equivalent. Quantum mechanically one thinks in terms of some initial,
and thus  in the Heisenberg picture, eternal state.

 The hope has frequently been expressed in the past that eventually
theorists will  hit upon a unique theory, 
with all coupling constants determined by
consistency  or symmetry considerations, and  with just one superselection
sector. Even given such a ``theory of everything'' (TOE), there remains the issue of boundary conditions
or initial state, as emphasized by Hawking \cite{Hawking}.
Recently, however, there has been a considerable decline in optimism 
and few now seem
to believe in a  
single TOE with a single superselection sector, and many refer to 
a ``landscape''
of theories. 

One approach to this perceived crisis in theoretical physics  
is to resort to ``anthropic''  considerations and invoke the idea that
there may indeed exist, in the Platonic sense,  an enormous number
of ``possible universes,'' of which only very few will allow the development
of sentient beings, and even fewer will allow sentient beings like ourselves.
Thus one is led to contemplate the ensemble of all possible universes,
sometimes referred to as a ``multiverse'' \cite{Rees}. 
This ensemble is sometimes  thought of non-Platonically 
\footnote{in an Aristotelian sense?}  as an ensemble of connected subsets 
of a much bigger physically existing  universe, referred to as 
a meta-universe \cite{Vilenkin}.

At this point it may be helpful to remark, lest the daunting task 
of thinking about and making more precise, the nebulous
notion of such a multiverse should not be thought entirely a 
problem for  theorists seeking credit for making predictions
about the world we see about us: that the observers  and experimenters
must also face up to that task when
assessing the reliability of their measurements or the
extent to which they can confirm theoretical predictions.
All such activities are essentially Bayesian in character \cite{Bayes}
and require some  notion of ``priors,'' that is, 
some sort of {\it a priori} measure
of the space of possibilities.      

This problem has been addressed, with admittedly only partial success,
in a previous  paper \cite{GibbonsHawkingStewart} 
where the multiverse, for concreteness,  was identified with
the set of classical histories of a minisuperspace cosmological model.
A  well-defined and natural  local measure on the space
of classical histories is easily  constructed, but unfortunately
the total  measure of all histories, even in this finite-dimensional
truncation of the full set of solutions of Einstein's equations,
is infinite. The problem was recently revisited
in \cite{GibbonsTurok}.     

In the present paper, we shall turn to the problem 
of finding a natural  measure on the space of coupling constants. 
Thus the multiverse in the present paper is a set of 
Lagrangians parametrized by a manifold $X$ 
or ``moduli space,''  whose coordinates
consist of   masses, mixing angles, coupling constants,
etc., and we wish to place a natural measure on this space.       
We hope this will be useful for anthropic considerations such as those
of \cite{Tegmark}, where $X \equiv S^1 $, the circle
parametrizing the  phase of the axion. In that case the issue of
a measure was trivial, but in more complicated cases such as we shall consider
in the present paper, the situation is more complicated.
We also hope that the work in this paper will help in clarifying
the notion of ``fine-tuning,''  which is so prevalent in phenomenological
discussions.

The structure of this paper is as follows. After introducing the notion of geometric probability and outlining the Kobayashi-Maskawa theory of CP violation in the standard model, we discuss metrics on $SU(3)$ and its quotients in Sec. \ref{Su3met}, starting with a left-invariant metric on $SU(3)$ which induces a metric on the flag manifold $SU(3)/U(1)^2$. We perform a Kaluza-Klein type reduction on the left phases and discuss different possible metrics on the double quotient $U(1)^2 \backslash SU(3) / U(1)^2$, the space of Kobayashi-Maskawa matrices. We also discuss the metric used by Ozsv\'ath and Sch\"ucking\cite{OzsvathSchucking}, and argue that it lacks a geometrical justification.

We then use all metrics we have discussed to compute statistics of the Jarlskog invariant $J$. While the measure on $SU(3)/U(1)^2$ is independent from the choice of left-invariant metric, the measure on the double quotient is nonunique. We find that in each case the standard deviation $\Delta J$ (with $\langle J\rangle =0$) is about three orders of magnitude greater than the experimentally observed value of $J$, which appears to be finely tuned. In Sec. \ref{fine}, we do a closer numerical analysis of the probability distribution of $|J|$ on the double quotient, using several possible choices for the measure. We quantify the statement that a magnitude of CP violation as small as observed appears unlikely.

In Sec. \ref{gaussian}, we take a different viewpoint: We now consider random distributions on the space of mass matrices in the standard model. We therefore need to find a measure on the space of $3\times 3$ Hermitian matrices. We find that the simplest choice which gives convergent integrals over this space, and expectation values for squared quark masses which reproduce the observed values, is a Gaussian weighting function with four free parameters, which can be chosen appropriately. We then find that the standard deviation $\Delta J$ is much smaller for this measure, making the observed value of $J$ appear typical. We conclude that with an appropriate distribution which respects the known quark mass hierarchy there is no need for fine-tuning in $J$.

We briefly discuss the case of neutrinos in Sec. \ref{neutrinos}, explaining the general theory, and the difference between Dirac and Majorana masses. We cannot give reliable predictions for $\Delta J$ for neutrinos, due to the absence of known values for their masses. 

\subsection{Geometric Probability}

The construction of appropriate measures over spaces 
of geometric objects goes back to the 18th century  cosmologist
Buffon and his celebrated needle problem
\cite{Buffon}. The reader may find a general account of the subject in
\cite{Santalo}.  The simplest case to consider is when 
the space of coupling constants  $X$ may be regarded
as a finite-dimensional homogeneous space with 
respect to some Lie group $G$ of symmetries,
and the  stabilizer or little group is  $H \subset G$. Thus $X=G/H$. 
If ${\rm dim} X=n$, 
our aim is to construct an $n$-form on $X$ which is invariant under the 
action of $G$.
In the case that $X=G$, as in the example of the axion circle, 
this is completely unambiguous.
We pick {\sl any} $n$-form at the unit element $e\in G$ and spread it 
over $G$ by left or right translation.
On a unimodular group, such as a compact group or a semisimple group, 
left or right translation will
give  identical results. The original $n$-form, being a top degree form,
is unique up to a multiple.
This multiple can be fixed by normalizing the total measure to unity.   
The normalized measure is therefore unique. 

We could, if we wished, construct the measure as the Riemannian volume 
element of {\it any}   left or right-invariant metric
on $G$. The result would be the same. In practice, a convenient procedure 
for calculating the measure
could be to construct an invariant metric on $X=G$ and then calculate 
its Riemannian volume element.
Often, the bi-invariant or Killing  metric is the most convenient  choice.

   In the case of a coset, $X=G/H$, the measure can again be taken to be 
any $n$-form at some arbitrarily
chosen point $x\in X$, which is then spread around using the group action.  
Since any $n$-form at $x$ will be 
$H$-invariant, the result is again unique and invariant under all 
the symmetries of the problem.  
Of course it is possible that one may express $X=G/H$ in more than one way. 
This could in principle
give rise to some discrete nonuniqueness, but in practice this seems not to be 
important.  

   Although the situation when coupling constants may be regarded as belonging
to a homogeneous space is quite satisfactory, it is often the case that 
coupling constants belong to an inhomogeneous space.  In particular, in the
case of ``mixing angles,'' they typically belong to a double coset, or
bi-quotient, of the form $H_1\backslash G/H_2$, where $H_1$ and $H_2$ are
(not necessarily identical) Lie subgroups of $G$.  The reason for this is 
that mixing angles relate two unitary bases for the same space of physical
states.  The two unitary bases may not be unique.  In particular, it is
often the case that the individual basis vectors can be multiplied by
arbitrary phases.  In this case, $H_1$ and $H_2$ may belong to $U(1)^k$, 
where $k$ is the number of states in the basis.  In the case of the 
Kobayashi-Maskawa matrix the states are quarks, and one basis diagonalizes
the strong Hamiltonian while the other basis diagonalizes the weak
interaction quantum numbers.

   A biquotient, or double coset, $H_1\backslash G/H_2$, is typically
{\it not} a homogeneous space. This is because the left action of $G$
will not in general commute with $H_1$, and similarly, the right action of
$G$ will not commute with $H_2$.  As a consequence, one cannot, in the
case of biquotients, use group invariance to construct an unambiguous
measure on the space of mixing angles.  Later in this paper, we shall 
explore in detail some available options, and the extent to which they
affect the probability distribution of mixing angles.

\subsection{CP Violation and the Jarlskog Invariant}
\label{cpviol}

  In this section we shall review the Kobayashi-Maskawa theory of 
CP violation in the quark sector of the standard model.

If $m$ and $m^\prime$ are the (Hermitian) mass matrices for the charge
${2 \over 3}$ and $-{1 \over 3} $ quarks, respectively, then 
there  exist  unitary matrices $U$ and $U^\prime$ 
such that
\be
UmU^\dagger = {\rm diag} (m_u,m_c,m_t)\,,\qquad
U^\prime m^\prime {U ^\prime}^\dagger = {\rm diag} (m_d,m_s,m_b)\,.
\ee
The Kobayashi-Maskawa matrix $V$ is defined by
\be
 V= U {U ^\prime}^\dagger\,.\label{ckmmatrix}
\ee
The normalized
mass eigenstates are only defined up to a phase, and changing these
phases changes the matrices $U$ and $U'$ according to
\be
U\longrightarrow P_L\, U\,,\qquad U'\longrightarrow P_R^\dagger\, U'\,.
\ee
Hence the Kobayashi-Maskawa matrix changes according to
\ben
V \rightarrow P_L V P_R\,,\label{Vtrans}
\een
where $P_L$ and $P_R$ are diagonal matrices belonging to $SU(3)$.
In other words, $P_L$ and $P_R$ may each be thought of as
belonging  to $T^2 \equiv U(1) \times U(1)$, the maximal
torus of $SU(3)$. Thus the four-dimensional
space of CP violating  parameters should be thought of as an element
of the double coset, or biquotient, $U(1)^2\backslash SU(3)/U(1)^2$, whereas the matrices $U$ and $U'$ should be thought of as elements of the left coset $U(1)^2\backslash SU(3)$.

In the discussion of geometric probability attempted in this paper, one could take the viewpoint that $U$ and $U'$ are the fundamental objects relevant in CP violation, which would lead to discussing distributions on $(U(1)^2\backslash SU(3))^2$. One can then use the fact that only $V=U {U'}^{\dagger}$ appears in the Kobayashi-Maskawa theory to reduce this to a distribution on a single $U(1)^2\backslash SU(3)$, as we shall see in Sec. \ref{flagman}. Alternatively, one considers $V$ as fundamental and considers the biquotient.

   Because the right action of $U(1)^2$ is free, the intermediate
coset $SU(3)/U(1)^2$ is a compact smooth homogeneous space
without boundary,
on which $SU(3)$ acts by left actions. In fact $SU(3)/U(1)^2$ is an example
of a flag manifold.  The maximal torus
$U(1)^2$ acts on the flag manifold
via left actions of $SU(3)$, but its action  on $SU(3)/U(1)^2$ is not
free, and as a consequence,
the biquotient $U(1)^2\backslash SU(3)/U(1)^2$ is not a smooth
compact manifold  without boundary.  Rather, it is
a stratified set whose boundary consists of components at  which either
or both of the left-acting $U(1)$ factors has fixed points.

   In the standard notation
\ben
V= \begin{pmatrix}
V_{ud} & V_{us} & V _{ub} \cr
V_{cd} & V_{cs} & V _{cb} \cr
V_{td} & V_{ts} & V _{tb} \cr
\end{pmatrix}\,,
\een
and it is customary to choose the phases so that
\be
V=\begin{pmatrix}
 1&0&0 \cr 0 & c_{23} & s_{23} \cr 0 & -s_{23} & c_{23}\end{pmatrix}
\begin{pmatrix} c _{13} &0& s_{13} e^{-i \delta} \cr 0 & 1  & 0 \cr -s_{13}
  e^{i\delta}
& 0 & c_{13}\end{pmatrix}
\begin{pmatrix} c_{12} &s_{12} &0\cr -s_{12}  & c_{12} & 0 \cr 0 &
0 & 1\end{pmatrix}\,,\label{Vc123}
\ee
where $s_{12}= \sin \theta_{12}$, $c_{12}=\cos\theta_{12}$, etc., and
the angles $\theta_{12}$, $\theta_{13}$, and $\theta_{23}$ are taken all
to lie in the first quadrant (\ie between 0 and $\ft12\pi$).

    One can take the angle $\delta$ as a measure of CP violation, but
its definition depends on the choice of phases.  Jarlskog 
\cite{Jarlskog:19851,Jarlskog:19852} introduced 
a formalism that eliminates this arbitrariness.  She defined a
Hermitian tracefree matrix $C$ by
\ben
\bigl [m,m^\prime \bigr] = \im C\,,
\label{comm}
\een
and took $\det C$ as a measure of CP violation.  She showed that
\ben
\det C= -2 T B J\,,
\label{determ} 
\een
where
\be
T=(m_t-m_u)(m_t-m_c)(m_c-m_u)\,,\qquad 
B=(m_b-m_d)(m_b-m_s)(m_s-m_d)\,,
\ee
and the Jarlskog invariant $J$ is given by
\ben
J= \Im \bigl (V_{11} \,V_{22}\, V_{12}^* \, V_{21}^* \bigr )
\,. \label{Jdef}
\een

   Despite appearances, $J$ is independent of the arbitrary phases.  In
other words, it is invariant under (\ref{Vtrans}).  In fact, it has
an extremely elegant geometrical interpretation.  Since $V$ is a unitary
matrix, its three rows and columns are orthogonal.  Thus, for example,
there are three relations of the form
\be
(V V^\dagger)_{12}= V_{11}\, V^*_{21} + V_{12}\, V^*_{22} +
   V_{13}\, V^*_{23}=0\,.
\ee
The three complex numbers $a=V_{11}\, V^*_{21}$,  $b=V_{12}\, V^*_{22}$, and
$c=V_{13}\, V^*_{23}$, satisfying $a+b+c=0$,  
may be thought of as the three sides of a {\it unitarity triangle}
in the complex plane.  The absolute value of $J$ is twice the area of this
triangle:
\be
|J|= |\Im(a b^*)| = |\Im(a c^*)| = |\Im( b c^*)|\,.
\ee
The effect of the transformation (\ref{Vtrans}) is to rotate this
triangle in the complex plane, but the area $\ft12|J|$ is unchanged.  Less
obviously, the same area results from taking either of the two other
possible inner products, $(V V^\dagger)_{13}=$ or $(V V^\dagger)_{23}=0$.
Thus $J$ is an invariant, and so it is well defined on the space of 
mixing angles.

  In terms of the standard parametrization (\ref{Vc123}), the Jarlskog
invariant is given by
\be
J= c_{12}\, c_{23}\, c_{13}^2\, s_{12}\, s_{23}\, s_{13}\, \sin\delta\,.
\label{Jexp1}
\ee

One could choose to take a different quantity as a measure of CP violation. Jarlskog \cite{jarlskog3} suggested appropriately normalizing the determinant (\ref{determ}) and using
\ben
a_{CP}=3\sqrt{6}\frac{\det C}{({\rm Tr }\,C^2)^{3/2}}\,,
\een
which takes values between $+1$ and $-1$ and is zero if and only if CP is conserved. Written out explicitly in terms of the quark masses and mixing angles, this is a complicated expression that we do not give here. As in the present paper the observed quark mass hierarchy is assumed, we shall not consider the case of coinciding quark masses, and we concentrate on $J$ as a measure of CP violation.

Another possible source of confusion is the assumption of general, not necessarily Hermitian, mass matrices. In this case the commutator (\ref{comm}) is replaced by
\ben
\bigl [mm^{\dagger},m^\prime {m^\prime}^{\dagger}\bigr] = \im {\bf C}
\een
in order for ${\bf C}$ to be Hermitian. The use of $C$ or ${\bf C}$ can lead to ambiguous ``orders of magnitude" estimates for CP violating processes, e.g. when discussing baryogenesis. We will assume that $m$ and $m'$ are Hermitian, and as our calculations only involve $J$ these considerations will not be relevant.

\section{Metrics on $SU(3)$ and its Quotients}\label{Su3met}

  A generic element $U$ of $SU(3)$ is conveniently parametrized by eight
real coordinates $(p,q,r,t,x,y,z,w)$, so that 
\ben
U= T_L\, W\, T_R\,,\label{Udef}
\een
where
\ben
T_L= e^{\ft{\im}{2}\, (3p-q)\lambda_3 + \ft{\im \sqrt3}{2}\, (p+q)\lambda_8}\,,
\qquad T_R= e^{\im t \lambda_3 + \im \sqrt3\, r\lambda_8}\,,
\een
and
\ben
W= e^{\im x \lambda_7}\, e^{-\im w \lambda_3}\, e^{\im y \lambda_5} \,
    e^{\im w \lambda_3}\, e^{\im z \lambda_2}\,,
\een
with
\be
0\le x\le\ft12\pi\,,\quad 0\le y\le \ft12\pi\,,\quad 0\le z\le \ft12\pi\,,
\qquad 0\le w \le 2\pi\,.
\ee
Here, we are using the standard Gell-Mann representation for the generators
of $SU(3)$:
\bea
&&\lambda_1=\begin{pmatrix} 0&1 &0\\ 
                          1&0&0\\
                          0&0&0 \end{pmatrix}\,,\qquad
\lambda_2=\begin{pmatrix} 0&-\im &0\\
                          \im &0&0\\
                          0&0&0 \end{pmatrix}\,,\qquad
\lambda_4=\begin{pmatrix} 0&0&1\\
                          0&0&0\\
                          1&0&0 \end{pmatrix}\,,\nn\\
&&\lambda_5=\begin{pmatrix} 0&0&-\im\\
                          0&0 &0\\
                          \im&0&0 \end{pmatrix}\,,\qquad
\lambda_6=\begin{pmatrix} 0&0&0\\
                          0&0&1\\
                          0&1&0 \end{pmatrix}\,,\qquad
\lambda_7=\begin{pmatrix} 0&0&0\\
                          0&0&-\im\\
                          0&\im &0 \end{pmatrix}\,,\nn\\
&&\lambda_3 =\begin{pmatrix} 1&0&0\\
                             0&-1&0\\
                             0&0&0 \end{pmatrix}\,,\qquad
\lambda_8 = \fft1{\sqrt3} \begin{pmatrix} 1&0&0\\
                             0&1&0\\
                             0&0&-2 \end{pmatrix}\,,
\eea
Explicitly, the matrices $W$, $T_L$, and $T_R$ are given by
\bea
W&=&\begin{pmatrix} c_y c_z & c_y s_z & e^{-\im w} s_y\cr 
    -c_x s_z - e^{\im w} s_x s_y c_z & c_x c_z - e^{\im w} s_x s_y s_z&
    s_x c_y\cr 
   s_x s_z - e^{\im w} c_x s_y c_z & - s_x c_z - e^{\im w} c_x s_y s_z &
    c_x c_y \end{pmatrix} \,,\\
T_L &=&\hbox{diag}\Big( e^{2\im p}, e^{-\im p+ \im q}, e^{-\im p -\im q}\Big)
\,,\\
T_R &=&\hbox{diag}\Big(e^{\im r + \im t}, e^{\im r - \im t},
    e^{-2\im r}\Big)\,,
\eea
where we use the notation $s_x=\sin x$, $c_x=\cos x$, etc.

   If we identify $W$ as the Kobayashi-Maskawa matrix $V$ in the standard
conventions (\ref{Vc123}), then
\be
x=\theta_{23}\,,\qquad y= \theta_{13}\,,\qquad z=\theta_{12}\,,\qquad
   w=\delta\,.
\ee

   If we define left-invariant one-forms $\sigma_a$ by
\ben
U^{-1}\, dU = \im\, \lambda_a\, \sigma_a\,,\label{sigmadef}
\een
then the general left-invariant metric on $SU(3)$ takes the form
\be
ds^2 = g_{ab}\, \sigma_a\, \sigma_b\,,\label{genmet}
\ee
where $g_{ab}$ is a constant symmetric matrix.

    For a general choice of the matrix $g_{ab}$, the metric admits 
no further isometries beyond the left action of $SU(3)$, which we denote
by $SU(3)_L$.  For special
choices of $g_{ab}$, however, the metric is additionally 
invariant under the right action of some subgroup $K$ of $SU(3)_R$.
The most symmetric such case, the bi-invariant or Killing metric 
for which $K$ is the full right-acting
$SU(3)_R$, arises if $g_{ab}$ is proportional to $\delta_{ab}$.
The various intermediate possibilities, of which there are five, 
are listed in \cite{coqjad}.  

    In the generic
case (\ie when $K$ is the identity), $28=36-8$ parameters are required
to specify the metric.  One of these parameters sets the overall scale.
For the intermediate cases there are correspondingly fewer parameters
\cite{coqjad}.  The bi-invariant metric has the smallest number, namely
just the overall scale.  In all cases, the invariant measure on the
group $SU(3)$ is the same and given by
\be
\mu = N\, \prod_a \sigma_a\,,
\ee
where $N$ is a constant normalization factor.  

   One of the intermediate cases given in \cite{coqjad} corresponds to
\be
ds^2= a^2 \,(\sigma_1^2 + \sigma_4^2 + \sigma_6^2) + b^2\, (\sigma_2^2 +
\sigma_5^2 + \sigma_7^2) + a^2\, (\sigma_3^2 + \sigma_8^2)\,.
\ee
This has the symmetry $SU(3)_L\times SO(3)_R$, where the $SO(3)_R$ is 
generated by $\lambda_2$, $\lambda_5$ and $\lambda_7$.  Remarkably, there
is a second Einstein metric in this class \cite{Jensen}, in addition to
the standard bi-invariant metric that arises when $a=b$.  The nonstandard
Einstein metric occurs when $b=a/\sqrt{11}$.

    However, the measure on the biquotient $U(1)^2\backslash SU(3)/U(1)^2$ 
of $SU(3)$ is not unique.  In particular, if one constructs the measure
from an invariant metric it will depend upon the metric that is used.
One of the cases enumerated in \cite{coqjad}, which is of particular 
interest for purposes, is when $K=U(1)\times U(1)$. One may denote
these $U(1)$ subgroups by $U(1)_3$ and $U(1)_8$, indicating that they
are generated by $\lambda_3$ and $\lambda_8$.  The possible 
$U(1)_3\times U(1)_8\times SU(3)_L$ invariant metrics on $SU(3)$ are
\be
ds^2 = \alpha\, (\sigma_1^2+\sigma_2^2) + \beta\, (\sigma_4^2+\sigma_5^2)
   + \gamma\, (\sigma_6^2+\sigma_7^2) + \delta_1 \, \sigma_3^2 +
  \delta_2\, \sigma_8^2 + 2\delta_3\, \sigma_3 \sigma_8\,.
\label{squashings}
\ee
The induced metric on the right coset $SU(3)/U(1)^2$ is then
given by
\be
ds^2 = \alpha\, (\sigma_1^2+\sigma_2^2) + \beta\, (\sigma_4^2+\sigma_5^2)
   + \gamma\, (\sigma_6^2+\sigma_7^2)\,.\label{flagsquash}
\ee
The normalized invariant measure on this coset is given by
\be
\mu = N\, \sigma_1\wedge \sigma_2\wedge \sigma_4\wedge\sigma_5\wedge
   \sigma_6\wedge \sigma_7\,.
\label{flagmeas}
\ee

   There is no similarly unique construction of a measure on the 
biquotient $U(1)^2\backslash SU(3)/U(1)^2$, because there is no
natural action of $SU(3)_L$ on it.  The reason for this is that the
$U(1)^2$ of the left quotienting is the maximal torus in $SU(3)_L$, and
so no other generators commute with it. 

    Locally, the biquotient $U(1)^2\backslash SU(3)/U(1)^2$ is a fiber
space whose fibers are orbits of $U(1)^2_L\times U(1)^2_R$,
whose action has fixed points.  The biquotient is therefore not a 
smooth manifold.  Nevertheless, any metric on $SU(3)$ will induce on
any local section a metric, and hence a Riemannian measure.  However, 
the metric and the measure will in general depend upon the choice of
section.  In the language of Kaluza-Klein theory, such metrics will
in general depend upon the choice of gauge.  

   One way to resolve this ambiguity is to project the initial metric on
$SU(3)$ orthogonally to the orbits of $U(1)^2_L\times U(1)^2_R$.  The resulting
Kaluza-Klein metric shall be discussed in Sec. \ref{KKbisec}.

\subsection{The flag manifold $SU(3)/U(1)^2$}\label{EKsec}

   A choice of metric on the coset 
$SU(3)/U(1)^2$ can give rise to different metrics on the
bi-quotient, depending on the choice of section we make.  In what 
follows, we shall illustrate this by choosing a natural metric on the
flag manifold $SU(3)/U(1)^2$ that is Einstein-K\"ahler. 

  There is a general construction showing that every quotient of a
compact Lie group $G$ by its maximal torus may be regarded as an
Einstein-K\"ahler manifold.  A physical application of this result
would be to the modulus space of vacua of a Yang-Mills theory with
Higgs in the adjoint.  We shall describe the special case of $G=SU(3)$,
following the construction described in \cite{Picken}.   This makes
use of the fact that
\be
SU(3)/U(1)^2 = SL(3,\C)/B\,,
\ee
where $B$ is the Borel subgroup of $SL(3,\C)$.  In other words,
we can express an $SU(3)$ matrix $U$ in the Iwasawa form
\ben
U= \begin{pmatrix}  1&0&0 \cr  -z_3 & 1&0\cr  -z_2 & z_1& 1 \end{pmatrix}
\begin{pmatrix} u&0&0 \cr  0& v& 0 \cr  0&0& \ft1{uv}\end{pmatrix}
 \begin{pmatrix} 1 & y_1 & y_2\cr    0 & 1& y_3\cr  0&0&1 \end{pmatrix} \,.
\een 
Substituting the expression for $U$ given in (\ref{Udef}), we find 
in particular that
\bea
  z_1 &=& - e^{-2\im q}\, \tan x\,,\nn\\
  z_2 &=& e^{-3\im p -\im q}\, (e^{\im w}\, \cos x\, \tan y -
                       \sin x\, \sec y\, \tan z)\,,\label{zeds}\\
z_3 &=&  e^{-3\im p +\im q}\, (e^{\im w}\, \sin x\, \tan y + 
                   \cos x\, \sec y\, \tan z)\,.\nn
\eea

   These expressions can be inverted to give the real coordinates
in terms of the $z^\alpha$:
\bea
\tan^2 x &=& | z_1|^2\,,\nn\\
\tan^2 y &=& \fft{|z_2- z_1\, z_3|^2}{1+|z_1|^2}\,,\nn\\
\tan^2 z &=&
  \fft{|z_3+\bar z_1\, z_2|^2}{1+ |z_1|^2 + |z_2-z_1\, z_3|^2}\,,\nn\\
e^{\im w} &=& \fft{(z_3 \tan x - \bar z_1 \, z_2) \tan z}{(z_3 + \bar z_1\,
        z_2\, \tan x)\sin y}\,,
\eea
with $p$ and $q$ then obtained using
\be
\fft{z_1}{\bar z_1} = e^{-4\im q}\,,\qquad
 \fft{z_2}{\bar z_2} = e^{-6\im p -2\im q + 2\im w}\,.
\ee

  As discussed in \cite{Picken}, the $z^\alpha$ can be viewed as 
complex holomorphic
coordinates on the flag manifold.  The K\"ahler function is given by
\ben
K= \log(1+|z_2|^2 + |z_3|^2) + \log(1+ |z_1|^2 + |z_2- z_1\, z_3|^2)\,.
\label{kfun}
\een
It is easy to check that the K\"ahler metric, given by
\ben
ds^2 = g_{\alpha\bar\beta} \, dz^\alpha\, d\bar z^{\bar\beta} \,,\qquad
  g_{\alpha\bar\beta} = \fft{\del^2 K}{\del z^\alpha\, \del\bar z^{\bar\beta}}
\,,\label{Kahler}
\een
has a determinant given by
\ben
\det(g_{\alpha\bar\beta}) = 2(1+|z_2|^2 + |z_3|^2)^{-2}\, 
                             (1+ |z_1|^2 + |z_2- z_1\, z_3|^2)^{-2}\,,
\een
which can therefore be written as
\ben
\det(g_{\alpha\bar\beta}) = 2 e^{-2K}\,.
\een
Thus $K$ satisfies the Monge-Amp\`ere equation, implying that the K\"ahler
metric $g_{\alpha\bar\beta}$ is Einstein.  (Since $R_{\alpha\bar\beta}=
 \del_\alpha\del_{\bar\beta} \log(\sqrt{g})$.)  

    Substituting (\ref{zeds})
into (\ref{Kahler}), one obtains the Einstein-K\"ahler metric on the
flag manifold written in terms of the real coordinates $(p,q,x,y,z,w)$.  
It is straightforward to verify directly that it satisfies
\ben
R_{ij} = 4 g_{ij}\,.
\een
In terms of the real coordinates, the K\"ahler function (\ref{kfun}) 
is given by
\ben
e^{-K} = \cos^2 x\, \cos^2 z\, \cos^4 y\,.
\een

   The Einstein-K\"ahler metric (\ref{Kahler}) is invariant under the 
left action of $SU(3)$, and in particular, under the $T^2$ action 
generated by $\del/\del p$ and $\del/\del q$.  From (\ref{zeds}), 
this action corresponds to phasing the complex coordinates $z^\alpha$ in
such a way as to leave the K\"ahler function (\ref{kfun}) invariant.  It
is possible, therefore, to perform a Kaluza-Klein reduction on the
two angles $p$ and $q$, to obtain a metric on the double coset
$U(1)^2\backslash SU(3)/U(1)^2$.  The resulting metric is
extremely complicated, and we shall not give it explicitly.  However, the metric we obtain is not the 
same as the one discussed in Sec. \ref{KKbisec}.  

   This difference is connected with the
fact that the Einstein-K\"ahler metric given by (\ref{kfun})
and (\ref{Kahler}) is not the ``round'' metric on $SU(3)/U(1)^2$,
but rather, it is a particular member of a one-parameter family of
homogeneous squashed metrics.  
(It corresponds to the only other
member of the family, other than the round metric, that is Einstein.)   
The Einstein-K\"ahler metric constructed in (\ref{Kahler}) 
is given, in terms of the left-invariant one-forms $\sigma_a$ defined in
(\ref{sigmadef}),  by
\ben
ds^2= \sigma  _1^2 + \sigma_2^2 + \sigma_6^2 + \sigma_7^2 + 
            2(\sigma_4^2+\sigma_5^2)\,.\label{ekmetric}
\een

   This illustrates the remarks we made previously about the ambiguity of
measures on bi-quotients.  The round metric on $SU(3)$ corresponds to
setting
\be
(\alpha,\beta,\gamma,\delta_1,\delta_2,\delta_3)=(1,1,1,1,1,0)
\ee
in (\ref{squashings}).   Kaluza-Klein reduction with respect to
$\del/\del r$ and $\del/\del t$ gives the round Einstein metric
corresponding to $\alpha=\beta=1$ in (\ref{flagsquash}).   The same
metric on the flag-manifold quotient would also arise for general values
of $\delta_1$, $\delta_2$, and $\delta_3$, as long as $\alpha=\beta=\gamma=1$.
If, on the other hand, 
\be
(\alpha,\beta,\gamma)=(1,1,2)\,,
\ee
for arbitrary $\delta_1$, $\delta_2$, and $\delta_3$, we obtain the
squashed Einstein-K\"ahler metric (\ref{Kahler}) on the flag manifold.

This construction, while not providing us with a ``simple" metric on the double quotient, has the virtue of being possible for any coset $SU(N)/U(1)^{N-1}$; we shall see in Sec. \ref{neutrinomix} that the case $N=6$ may be of relevance to neutrinos.

\subsection{Kaluza-Klein reduction of the bi-invariant metric}\label{KKbisec}

   Here, we start with the bi-invariant metric on $SU(3)$,
\be
ds^2=  \ft12 \hbox{Tr }dU\,dU^{\dagger}= \sigma_a^2\,.\label{su3met0}
\ee
In terms of the coordinates $(p,q,r,t,x,y,z,w)$, it
is given by
\bea
ds^2 &=& 3dp^2+ dq^2 + 3 dr^2 + dt^2 + \ft32(3\cos 2y -1) dp dr +
3 \cos^2 y\, (\cos2 z\, dp dt + \cos 2x\, dq dr )\nn\\
&&+
\ft12\{\cos 2x \, \cos 2z\,
   (\cos 2y - 3) +4 \sin 2x\, \sin 2z\, \sin y\,
  \cos w\}\, dq dt\nn\\
&& - \sin^2 y\, (3dp -3 dr + \cos2x\, dq -\cos2z\, dt)dw 
+2 \sin y\, \sin w\, (\sin 2z\, dt dx - \sin 2x\, dq dz)\nn\\
&&+ dx^2+dy^2+dz^2 + \sin^2 y\, dw^2 + 2 \sin y\, \cos w\, dx dz\,.
\label{su3met}
\eea
As expected, this metric on $SU(3)$ does not depend on 
$p,q,r$ and $t$, which are the arbitrary quark phases appearing 
in the Kobayashi-Maskawa matrix.

   It is perhaps worth remarking here that the first expression in 
(\ref{su3met0}) is well defined for any complex matrices $U$, unitary or not.
For general complex matrices, it defines a flat metric on the space of
matrix elements, which may be identified with $\C^9\equiv 
 {\Bbb E}^{18}$, the 18-dimensional Euclidean space.\footnote{Obviously,
for $k\times k$ matrices, $\C^9$ is replaced by $\C^{k^2}$.}  $SU(3)$
may be regarded as a real eight-dimensional submanifold of 
$ {\Bbb E}^{18}$, defined by the nine real unitary constraints $U U^\dagger=1$
together with the one real unimodularity constraint $\det U=1$. The
bi-invariant metric on $SU(3)$ is the induced metric on this 
submanifold.

   One approach to placing a measure on mixing angles would be to give a
uniform measure on the unconstrained mixing angles, and then to obtain a
measure on the mixing angles by implementing the unitarity and unimodularity
conditions.  The left-invariant measure on $SU(3)$ is unique up to
a scale.  Thus, any construction which respects $SU(3)$ invariance will
result in a measure which is a constant multiple of the Riemannian
measure constructed from the bi-invariant metric.

   Writing (\ref{su3met}) in the standard Kaluza-Klein form,
\be
ds^2 = h_{ij}(x)\, (dy^i + A^i_\mu(x) dx^\mu)(dy^j + A^j_\nu(x) dx^\nu)
    + \tilde g_{\mu\nu}(x) dx^\mu\, dx^\nu\,,
\ee
where $y^i=(p,q,r,t)$ and $x^\mu=(x,y,z,w)$. The metric on the
bi-quotient is then given by
\be
d\tilde s^2= \tilde g_{\mu\nu}(x) dx^\mu\, dx^\nu\,.\label{biqmet}
\ee

   The metric (\ref{biqmet}) is once again rather complicated, and we shall not
present it explicitly since we really only wish to calculate the Riemannian 
measure
\be
\mu = \sqrt{\tilde g}\, dx dy dz dw\,.
\ee
Noting that $\det g = \det h\, \det \tilde g$, and that
\be
  \det g= \ft{27}{4} \sin^2 2x\, \sin^2 y\, \sin^2 2z\, \cos^6 y\,,
\ee
we find that, after extracting an unimportant overall constant factor,
\be
\det\tilde{g}=  \sin^2 2x\, \sin^2 2z\, \sin^2 y\, \cos^4 y/F
\,,\label{messy}
\ee
where
\bea
F&=& (\sin^2 2x+ \sin^2 2z)\sin^2y +\ft18 (5\cos 2y-3)\sin^2 2x \,\sin^2 2z
\nn\\
&& +\ft12 \sin 4x\, \sin 4z\, \sin^3 y\, \cos w 
+\ft18 (3\cos 2y -5)\sin^2 2x\, \sin^2 2z\, \sin^2 y\, \cos^2 w
\,.\label{bimeas}
\eea

   Note that one can alternatively obtain the four-dimensional metric
on the bi-quotient by means of a $T^2$ Kaluza-Klein reduction of the
round flag-manifold metric
\be
ds^2 = \sigma_1^2 + \sigma_2^2 + \sigma_4^2+\sigma_5^2 +\sigma_6^2 +
\sigma_7^2\,,\label{roundflag}
\ee
which differs from the one used in Sec. \ref{EKsec}.

   We have calculated the Riemannian metrics of the two different 
metrics on the bi-quotients, and confirmed that these two four-dimensional
measures are indeed different.  Later, we shall demonstrate the dependence
of the mean-square value of the Jarlskog invariant on the choice of
squashing.

\subsection{Squashed Kaluza-Klein metrics}\label{KKsqsec}

   As we noted earlier, not only is the four-dimensional
double-coset {\it metric} obtained by Kaluza-Klein reduction nonunique, but 
also the associated {\it measure} is nonunique.  In Sec. \ref{KKbisec},
we constructed the measure that follows from the $T^4$ 
Kaluza-Klein reduction of
the bi-invariant $SU(3)$ metric to four dimensions, or,  equivalently, 
the $T^2$ Kaluza-Klein reduction of the round six-dimensional 
flag metric (\ref{roundflag}).  Here,
we present the more general result for the measure on the double coset
that is obtained by Kaluza-Klein reducing a one-parameter family of squashed 
flag metrics.  Specifically, we take as our starting point the flag metrics
\be
ds^2 = \sigma_1^2+\sigma_2^2 + \sigma_6^2 + \sigma_7^2 + 
  \beta\, (\sigma_4^2+\sigma_5^2)\,.
\ee

    After Kaluza-Klein reduction, we find that the determinant of the four-dimensional metric $\tilde g_{\mu\nu}$ is given,
after again extracting an unimportant overall constant factor, by
\be
\det \tilde g = \sin^2 2x\, \sin^2 2z\, \sin^2 y\, \cos^4y/F\,,\label{sqmeas0}
\ee
where
\bea
F&=& (\sin^2 2x+ \sin^2 2z)\sin^2y +\ft18 (5\cos 2y-3)\sin^2 2x \,\sin^2 2z
   + \ft1{16} (\beta-1)^2 \,\sin^2 2x\, \sin^2 2y\, \sin^2 2z \nn\\
&& + (\beta-1) \Big[4 \sin^2 y \sin^2 z \cos^4 z \nn\\
&&\quad +\sin^2 2x\,  
                 \Big\{\cos^2 z\, \cos^2 2z -\ft14 \cos 2z\, \sin^2 2y -
  \cos^2 y\, \cos^4 z\, [\cos^2 y-(3+\sin^2y)\sin^2 z]\Big\}\Big]\nn\\
&& +\sin 4x\, \sin y\, \sin 2z\, \cos w\,\Big[ \cos 2z\, \sin^2 y +
  (\beta-1) (\sin^2 y \, \cos^2 z- \sin^2 z) \cos^2 z\Big]\label{sqmeas}\\
&&+\ft18 \sin^2 y\, \sin^2 2x\, \sin^2 2z\, \cos^2 w\, \Big[
  3\cos 2y-5 - 4(\beta-1) (\cos 2z + \sin^2 y)-2(\beta-1)^2\, \cos^2 y\Big]
\,.\nn
\eea
Note that this expression reduces to (\ref{bimeas}) if
$\beta=1$, which is the special case of the reduction of the round
flag metric. The nontrivial dependence of (\ref{sqmeas}) on the
squashing parameter $\beta$ shows that the measure $\sqrt{\tilde g}\, dx dy
dz dw$ on the double coset is also nontrivially dependent on the
choice of squashing.

\subsection{Ozsv\'ath-Sch\"ucking metric}

    The previous calculations give rise to rather complicated formulae.
It is striking, therefore, that the metric obtained by Ozsv\'ath and
Sch\"ucking \cite{OzsvathSchucking} is so much simpler.  
Their choice of section 
consists of simply setting $p=q=r=t=0$ in the metric (\ref{su3met}).  This
results in the metric 
\ben
ds ^2 =   dx^2 + dy^2 + d z^2
 + 2 \sin y \,\cos w \, dx dz + \sin ^2 y dw ^2 \,.
\label{metric}
\een
This metric is manifestly invariant under translating the coordinates
$x$ and $z$.  Remarkably, there is one further commuting Killing vector.
If one defines new coordinates $u$ and $v$ by
\ben
(\sin y\, \cos w, \sin y\, \sin w, \cos y)= (\cos u, \sin u\, \cos v,
   \sin u\, \sin v)\,,\label{embed}
\een
then the metric (\ref{metric}) takes the form
\ben
ds ^2 = du^2 + dx^2 + dz^2 +2 \cos u \, dx dz + \sin ^2 u\,  d v ^2 \,,
\label{ozsvath}
\een 
which has the three commuting Killing vectors $\del/\del x$, 
$\del/\del z$ and $\del/\del v$.

   Geometrically, we can understand this if we note that the metric
(\ref{metric}) may recast as
\be
ds^2 = (dx + \sin y\, \cos w\, dz)^2 + (1-\sin^2y\, \cos^2 w)\, dz^2+
  dy^2 + \sin^2 y\, dw^2\,,
\ee
which exhibits it as a $T^2$ fibration (having coordinates $x$ and $z$)
over a round hemisphere (having coordinates $y$ and $w$).  (The 
colatitude $y$ lies in the interval $0\le y\le \ft12\pi$.)  The hemisphere
can be embedded isometrically into Euclidean three-space as 
$(\sin y\, \cos w, \sin y\, \sin w, \cos y)$. Equation (\ref{embed})
then gives a different embedding such that while $\del/\del w$ generates
a rotation around the first axis, $\del/\del v$ generates a rotation 
around the third axis.  The projection along the third axis is given 
by $\cos y$, while the projection along the first axis is given by
$\cos u$.  

   Note that the extra Killing vector $\del/\del v$ is purely local,
since rotations about the first axis do not preserve the hemisphere.

   The metric (\ref{ozsvath}) can be recast in the form
\be
ds^2 = \sin^2 u\, dv^2 + (dz+ \cos u\, dx)^2 + du^2 + \sin^2 u\, dv^2\,,
\ee
which is locally of the form of a $U(1)$ fibration (with coordinate $v$)
over $S^3$.

   The un-normalized measure is given by the remarkably simple
formula
\bea
\mu &=& \sin y\, (1-\sin^2 y\, \cos^2 w)^{1/2} \, dx dy dz dw\,,\nn\\
&=& \sin^2 u\, du dv dx dz\,.\label{ozscmeasure}
\eea

   Despite its appealing simplicity, the Ozsv\'ath-Sch\"ucking construction
lacks a geometrical justification and introduces a spurious 
$U(1)^3$ symmetry into the problem.  A simpler example, which makes this
clear, is provided by considering the lower-dimensional example
of quotients of $SU(2)$.  The bi-invariant metric
on $SU(2)$ is
\be
ds^2 = (d\psi+\cos\theta\, d\phi)^2 + d\theta^2 + \sin^2\theta\, d\phi^2\,,
\ee
where $\del/\del\phi$ generates $U(1)_L$ and $\del/\del\psi$ generates
$U(1)_R$.   Projecting the metric orthogonally to the orbits of 
right translations, {\it \`a la} Kaluza-Klein, gives the round metric
\be
ds^2 = d\theta^2 + \sin^2\theta\, d\phi^2\label{s2met}
\ee
on $S^2$.  By contrast, simply setting $d\psi=0$ (the analog of
the construction of Ozsv\'ath and Sch\"ucking) instead gives the
flat metric 
\be
ds^2 = d\theta^2 + d\phi^2\,.\label{flatmet}
\ee
The round metric (\ref{s2met}) is invariant under $SO(3)$.  The flat
metric (\ref{flatmet}) appears to be invariant under the Euclidean group,
with $\del/\del\theta$ and $\del/\del\phi$ having the appearance of
translations,
but these are only local symmetries since $\phi$ is a 
periodic coordinate and $\theta$ lies in an interval.

The example of $SU(2)$ also illustrates the difference between taking the flag-manifold measure and the Kaluza-Klein measure on a biquotient. The biquotient $U(1)\backslash SU(2)/U(1)=U(1)\backslash S^2$ is just an interval. Its metric becomes, after performing another Kaluza-Klein reduction of (\ref{s2met}),
\ben
ds^2=d\theta^2\,.
\een
The measure would be $d\theta$, and not $d\theta\sin\theta$ as obtained by integrating a function $f(\theta)$ over the coordinate $\phi$. It is apparent from this simple example that there are inequivalent ways of calculating integrals of a function on a right quotient that is invariant under the left group action; namely, one can {\it either} reduce the metric to obtain a measure on the double quotient {\it or} take the measure on the single quotient and integrate out the left phases.

\section{Statistics of the Jarlskog Invariant $J$}

Expressed in terms of the coordinates $(x,y,z,w)$, the Jarlskog invariant
(\ref{Jexp1}) is given by
\ben
J= \ft14 \sin2x\,  \sin2z\, \sin y\, \cos^2 y\, \sin w\,.
\label{Jdef2}
\een 
The average of a function $f$ on a space with metric $\tilde g_{\mu\nu}$ is 
defined by
\be
\langle f \rangle = \fft{\int f \sqrt{\tilde g}\, dx \,dy\, dz\, dw}{
                        \int \sqrt{\tilde g}\, dx\, dy\, dz\, dw}\,.
\ee
   The experimental value of the Jarlskog invariant $J$ of the 
Kobayashi-Maskawa matrix
is
\be
J=3.08_{-0.18}^{+0.16}\times 10^{-5}\,,\label{Jckmval}
\ee
which is very small compared with its maximum value
\be 
J_{\rm max} = \fft1{6\sqrt 3} \approx 0.0962\,.
\ee

In the following 
subsections \ref{flagman} to \ref{uniformsec}, we calculate the moments
of $J$ for the Kobayashi-Maskawa 
matrix, using the various measures we have introduced,
and compare them with the experimental value. We will see that the average values one obtains are rather insensitive to the choice of measure.

In subsection 
\ref{flagman},
we start with $SU(3)$-invariant measures on the flag manifold $SU(3)/U(1)^2$.
In this case, as we have already noted, there is an unambiguous 
$SU(3)$-invariant measure. 

\subsection{The flag-manifold measure}
\label{flagman}

   In Sec. \ref{cpviol}, we saw that the Kobayashi-Maskawa
matrix 
is an element of the four-dimensional bi-quotient
$U(1)^2\backslash SU(3)/U(1)^2$, which is however composed of two elements of $U(1)^2\backslash SU(3)$. This left quotient is, just like the right
quotient $SU(3)/U(1)^2$, the flag manifold.  

   Since the Jarlskog invariant (\ref{Jdef}), or (\ref{Jdef2}), is independent of all
phasing angles, the averaging of its moments over
the flag manifold will give the same results regardless of whether
one constructs the manifold as the left quotient or the right quotient of 
$SU(3)$ by $U(1)^2$.  This is convenient because we have already presented
detailed results for the metrics on the right cosets $SU(3)/U(1)^2$. 

   A straightforward calculation shows that for the general class
of $SU(3)$-invariant flag metrics (\ref{flagsquash}),
\ben
\sqrt{g} = 3 \alpha\beta\gamma\, \sin 2x\, \sin 2z\, \sin y\, \cos^3 y\,.
\label{flagmanifold}
\een
Since an overall constant factor in the measure cancels out in the 
normalized averaging process, we see therefore that in contradistinction
to the situation for the double coset $U(1)^2\backslash SU(3)/U(1)^2$,
the natural measure on the flag manifold is unique. 

   One might think of using the Cartesian product of two flag manifolds for calculating the moments of the Jarlskog invariant for the Kobayashi-Maskawa 
matrix. In fact, one could equally well use $SU(3)$ instead of the flag manifold since neither the bi-invariant measure nor $J$ depend on the $U(1)^2$ angles. The natural measure on $SU(3)\times SU(3)$, induced from (\ref{su3met0}), is
\be
\mu = N\, \sigma_1\wedge \sigma_2\wedge \sigma_3 \wedge \sigma_4\wedge\sigma_5\wedge
   \sigma_6\wedge \sigma_7 \wedge \sigma_8 \wedge \sigma'_1\wedge \sigma'_2\wedge \sigma'_3 \wedge \sigma'_4\wedge\sigma'_5\wedge
   \sigma'_6\wedge \sigma'_7\wedge \sigma'_8\,.
\label{prodmeasure}
\ee

Since it is only $V=U{U'}^{\dagger}$ that enters into the CP violating parameters, one could consider $U$ and $V$ as independent variables, \ie write $U'=V^{\dagger}U$ for some matrix $V$. Then the Maurer-Cartan form on the second $SU(3)$ is
\ben
\im\, \lambda_a \sigma_a' \equiv {U'}^{\dagger}dU'= U^{\dagger} dU - U^{\dagger} (dV \, V^{\dagger}) U \, ,
\een
which gives $\sigma_a'=\sigma_a - h_{ab}\tau_b$, where $\tau_b$ are right-invariant forms on $SU(3)$ in terms of $V$ coordinates and $h_{ab}$ only depends on the $U$ coordinates. The measure (\ref{prodmeasure}), expressed in terms of $V$ and $U$ coordinates, is thus a product of a function of the $U$ coordinates and the natural measure in $V$ coordinates (left- and right-invariant forms on $SU(3)$ give the same measure). Integration over the $U$ coordinates then just gives an irrelevant constant, and one is left with the measure (\ref{flagmanifold}) on the space of $V$ matrices. This justifies the use of (\ref{flagmanifold}) instead of the more complicated constructions obtained by reducing to the double quotient, and we will regard (\ref{flagmanifold}) as the most natural choice of measure on the parameter space.

    For the measure (\ref{flagmanifold}) the evaluation of the necessary integrals is very simple
and we find that all odd powers of $J$ average to zero, and 
\ben
\langle J^2 \rangle = \fft1{720} \approx 1.389 \times 10^{-3}\,,\qquad
\langle J^4 \rangle =  \fft1{201600} \approx 4.960 \times 10^{-5}\,.
\label{flagmoments}
\een
Thus we find that $\Delta J$ for the Jarlskog invariant is given by
\be
\Delta J = \fft1{12\sqrt 5} \approx 0.0373\,,
\ee
which is about three orders of magnitude larger than the 
experimental value (\ref{Jckmval}).

\subsection{The Kaluza-Klein measure from the bi-invariant metric}
\label{biKKJsec}
  
For the metric on the biquotient discussed in Sec. \ref{KKbisec},
the expression for the measure is too complicated to allow us to 
perform the integrations analytically.  Using numerical integration,
we find that
\be
\langle J^2\rangle \approx 1.1161\times 10^{-3}\,,\qquad 
\langle J^4 \rangle \approx 3.750 \times 10^{-6}\,,\label{biKKmoments}
\ee
with the odd powers of $J$ again averaging to zero.
Thus we find
\be
\Delta J \approx \sqrt{\langle J^2 \rangle} \approx 3.341\times 10^{-2}\,,
\label{DJround}
\ee
which is very close to the previous result.

Naively, one might have thought that since $J$ is independent of all the $U(1)$
phases, the results would be the same whether one averaged over the
space $U(1)^2\backslash SU(3)/U(1)^2$, or else the
flag manifold $SU(3)/U(1)^2$.  Of course we know that this is not in fact
correct, since, as we have seen, the measure for the biquotient depends
nontrivially on the squashing parameters $\alpha$, $\beta$, and $\gamma$ in 
(\ref{flagsquash}) while the measure on the single quotient does not. Nevertheless, it is interesting to compare the expressions (\ref{flagmoments})
for $\langle J^2\rangle$ and $\langle J^4\rangle$ with the ones we obtained
in (\ref{biKKmoments}) for the biquotient averaging.  They are in fact
quite similar, although the values are larger in (\ref{flagmoments}) than
in (\ref{biKKmoments}).  (We will see in the following subsection that,
among the more general class of squashed biquotient measures, the 
values of $\langle J^2\rangle$ and $\langle J^4\rangle$ seem to be 
maximized by the round case (\ref{biKKmoments}).)

\subsection{The Kaluza-Klein measure from squashed metrics}\label{sqKKmeassec}

   We can repeat the calculations of Sec. \ref{biKKJsec} using the 
measure given by (\ref{sqmeas0}) and (\ref{sqmeas}) for the one-parameter
family of squashed Kaluza-Klein metrics.  In view of the complexity of
the measure, we must again resort to numerical integration.  

   The case
where the squashing parameter is chosen to be $\beta=2$ is of particular
interest, since this corresponds to the second Einstein metric on
the flag manifold, \ie the one associated with the Einstein-K\"ahler 
metric we discussed in Sec. \ref{EKsec}.  For this choice, we find
\be
\langle J^2\rangle \approx 1.1012\times 10^{-3}\,,\qquad
\langle J^4 \rangle \approx 3.678 \times 10^{-6}\,,\label{beta2KKmoments}
\ee
with the odd powers of $J$ averaging to zero.
Thus we find
\be
\Delta J \approx \sqrt{\langle J^2 \rangle} \approx 3.318\times 10^{-2}\,,
\ee
This is smaller than the value of $\Delta J$ we obtained in (\ref{DJround})
for the averaging over the Kaluza-Klein reduction of the bi-invariant
metric, but only by about 0.7\%.  This does not bring it significantly
closer to the experimental value of $J$, given in (\ref{Jckmval}).

    One might wonder whether, for some sufficiently large or small
choice for the squashing parameter $\beta$, it might be possible to 
obtain a result for $\Delta J$ that was comparable with the experimentally
observed value.  In fact, it appears that $\langle J^2\rangle$ is a 
rather slowly varying function of $\beta$.  The value of
$\langle J^2\rangle$ appears to be maximized by the choice $\beta=1$, and to
fall off monotonically in both directions as $\beta$ is taken to zero or to
infinity. 

   For example, if we choose $\beta=\ft12$ we find
\be
\langle J^2\rangle\approx 1.103 \times 10^{-3}\,,\qquad
\Delta J \approx 3.321 \times 10^{-2}\,,
\ee
while if we take $\beta\rightarrow 0$ we find 
\be 
\langle J^2\rangle\approx 8.097 \times 10^{-4}\,,\qquad
\Delta J \approx 2.846 \times 10^{-2}\,.
\ee

  Taking $\beta=1000$, we find
\be
\langle J^2\rangle\approx 4.298 \times 10^{-4}\,,\qquad
\Delta J \approx 2.073 \times 10^{-2}\,,
\ee
while for $\beta = 10^6$ we find
\be
\langle J^2\rangle\approx 1.958 \times 10^{-4}\,,\qquad
\Delta J \approx 1.399 \times 10^{-2}\,.
\ee
Even quite extreme values for the squashing parameter only bring about
small reductions in $\Delta J$.

\subsection{The Ozsv\'ath-Sch\"ucking measure}

Using the Ozsv\'ath-Sch\"ucking measure (\ref{ozscmeasure}),
we find that  $\langle J\rangle =0$ and
\ben
\langle J^2 \rangle = 35\times 2^{-16} \approx  5.341 
\, \times 10^{-4} \,,\qquad  \langle J^4\rangle = 
27027\times 2^{-34}\approx  1.573 \,\times 10^{-6} \,,
\een
and that the standard deviation is 
\ben
\Delta J^2 = \sqrt{\langle J^4 \rangle - \langle J^2 \rangle ^2 } 
= \sqrt{22127}\times 2^{-17} \approx 1.135\, \times 10^{-3}\,, 
\een
and
\ben
\Delta J= \sqrt{ \langle J^2 \rangle  }=\fft{\sqrt{35}}{256}\approx
 2.311 \, \times 10^{-2} \,.
\een
Again, the results are rather similar to the previous cases.

\subsection{The uniform measure}\label{uniformsec}

   Assuming a uniform distribution over the angles, and hence treating
the double coset as a flat four-dimensional manifold so that the
measure is simply $\mu=1$, would give
\be
\langle J \rangle =0,\quad \langle J^2 \rangle = \fft{1}{2048}
  \approx 4.883\times 10^{-4}\,,
\qquad \langle J^4\rangle =189\times 2^{-27} \approx 1.408\times 10^{-6}\,.
\ee
Hence, this simplest possible choice gives
\ben
\Delta J \approx 2.210 \, \times 10^{-2} \,.
\een

\section{Fine-tuning of $J$}
\label{fine}

In the previous section we saw that different measures on the space of mixing angles all seem to lead to expectation values for $J$ which are about three orders of magnitude larger than the observed value. The value for $J$ that we observe hence appears to be finely tuned. In this section we shall do a closer, mainly numerical, analysis of the fine-tuning involved. We compare results obtained by taking the $SU(3)$-invariant and Kaluza-Klein measures, which seem natural from a geometric perspective, with a uniform distribution which is just the simplest possible choice.

\subsection{Probability distribution of $J$}
\label{probdist}
The observed value for the Jarlskog invariant $J$ is
\ben
J\approx 10^{-4.51}\approx e^{-10.39}\,.
\een
In order to obtain a probability distribution for $J$ we have used {\it Mathematica} to numerically compute integrals of the form
\ben
\int \sqrt{\tilde{g}}\,dx\,dy\,dz\,dw\;\theta(a-|J|)\;\theta(|J|-b)\equiv P(b \le|J|\le a)\cdot \int \sqrt{\tilde{g}}\,dx\,dy\,dz\,dw
\een
using Monte Carlo methods. The $SU(3)$-invariant flag-manifold measure and the Kaluza-Klein measure disfavor small values of $J$ more strongly than a uniform distribution would. For example, we obtain
\ben
P_{{\rm flag}}(|J|\le 10^{-4})\approx 0.25 \%\,,\quad P_{{\rm KK}}(|J|\le 10^{-4})\approx 0.44 \%\,.
\een
Taking a uniform distribution $\sqrt{\tilde{g}}\equiv 1$, we get
\ben
P_{{\rm unif}}(|J|\le 10^{-4})\approx 7\%\,.
\een
\begin{figure}[htp]
\centering
\includegraphics[scale=0.9]{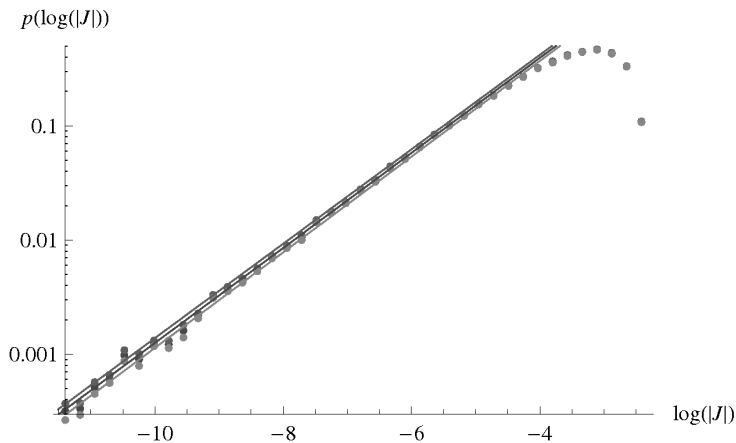}
\caption{{\small Probability distribution for $\log|J|$ using the $SU(3)$-invariant flag measure, with fit to $p(|J|)\propto |J|^{\lambda}$.}}
\end{figure}
\\The degree of fine-tuning required to reproduce a very small $J$ is considerably higher if one uses the measure induced by a $SU(3)$-invariant flag metric or the Kaluza-Klein metric, maybe contrary to what one might expect. Values of $J$ close to its maximal value of $\frac{1}{6\sqrt{3}}\approx 0.0962$ are disfavored in both cases. Therefore we have used a logarithmic scale for $|J|$. 
\begin{figure}[htp]
\centering
\includegraphics[scale=0.9]{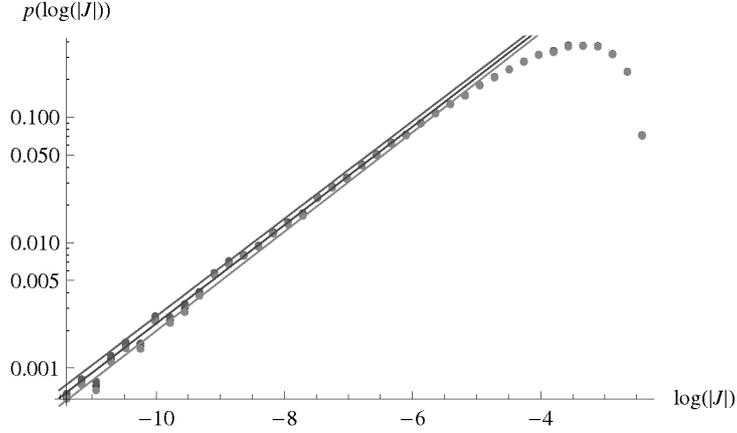}
\caption{{\small Probability distribution for $\log|J|$ using the Kaluza-Klein measure, with fit to $p(|J|)\propto |J|^{\lambda}$.}}
\end{figure}
\\In all three cases the numerical results for small $|J|$ are well approximated by a power law of the form $p(|J|)=\alpha\cdot |J|^{\lambda}$ for the probability density of $|J|$. The logarithmic graphs show $p(\log|J|)\propto |J|^{\lambda+1}$. For the $SU(3)$-invariant flag measure (Fig. 1), the best fit to the data in the region below $|J|=10^{-2.3}$ or $\log|J|=-5.3$ is
\ben
\lambda_{{\rm flag}}=-0.042(\pm 0.006)\,,\quad\alpha_{{\rm flag}}=18.1(\pm 0.7)\,;
\een
for the Kaluza-Klein measure (Fig. 2) we fitted the data in the region below $|J|=10^{-2.7}$ or $\log|J|= -6.2$ and obtained
\ben
\lambda_{{\rm KK}}=-0.097(\pm 0.008)\,,\quad\alpha_{{\rm KK}}=18.9(\pm 1.0)\,;
\een
\begin{figure}[htp]
\centering
\includegraphics[scale=0.9]{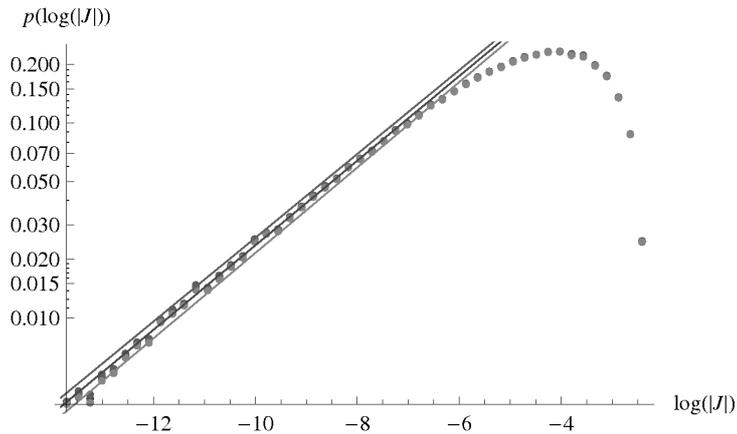}
\caption{{\small Probability distribution for $\log|J|$ using a uniform distribution, with fit to $p(|J|)\propto |J|^{\lambda}$.}}
\end{figure}
finally for the uniform measure (Fig. 3), the best fit to the data in the region below $|J|=10^{-3.4}$ or $\log|J|= -7.8$ is
\ben
\lambda_{{\rm unif}}=-0.500(\pm 0.005)\,,\quad\alpha_{{\rm unif}}=3.51(\pm 0.15)\,.
\een

\subsection{Wolfenstein parametrization}
\label{wolfenstein}
A different parametrization of the Kobayashi-Maskawa matrix which is frequently used was introduced by Wolfenstein and is based on the experimentally observed hierarchy
\ben
y\ll x\ll z\ll 1
\een
in the mixing angles. One rewrites \cite{Wolfenstein} 
\ben
\sin z=\lambda\,,\;\sin x=A\lambda^2\,,\;\sin y e^{-\im w}=A\lambda^3 (\rho-\im\eta)
\een
and treats $\lambda$ as a small parameter while $A,\rho$, and $\eta$ are supposed to be parameters of order unity. In the modern literature one also frequently uses $\bar\rho,\bar\eta$ instead of $\rho$ and $\eta$ because then the combination $\bar\rho + i\bar\eta$ is independent of the phase convention in the 
Kobayashi-Maskawa matrix \cite{particledb}. These parameters are defined by
\ben
\rho=\sqrt{\frac{1-A^2 \lambda^4}{1-\lambda^2}}\frac{\bar\rho-A^2\lambda^4(\bar\rho^2+\bar\eta^2)}{(1-A^2\lambda^4 \bar\rho)^2+A^4\lambda^8\bar\eta^2}\,,\quad \eta=\sqrt{\frac{1-A^2 \lambda^4}{1-\lambda^2}}\frac{\bar\eta}{(1-A^2\lambda^4 \bar\rho)^2+A^4\lambda^8\bar\eta^2}\,.
\een
The experimental values for $\lambda, A, \bar\rho, \bar\eta$ are \cite{particledb}\footnote{Note that only even powers of $\lambda$ appear in all expansions, so that it is $\lambda^2\approx 0.05$ which is the small parameter.}
\ben
\lambda=0.2272\pm 0.0010,\; A=0.818_{-0.017}^{+0.007},\; 
\bar\rho=0.221_{-0.028}^{+0.064},\;\bar\eta=0.340_{-0.045}^{+0.017}\,.
\een
One viewpoint on the Wolfenstein parametrization is that it is adapted 
to the values for the Kobayashi-Maskawa matrix entries that we observe 
and has no deeper significance; but often the viewpoint is expressed that 
this parametrization expresses some kind of ``natural hierarchy" in the 
mixing angles coming from physics beyond the standard model (see e.g. 
\cite{biggi}). Treating the other parameters as ``naturally of order unity" 
reduces our calculations to a one-dimensional problem as everything is only 
expanded in terms of $\lambda$. We find that the $SU(3)$-invariant measure on the flag 
manifold is now, to leading order in $\lambda$,
\ben
\left|\frac{\partial(x,y,z,w)}{\partial(\lambda,A,\bar\rho,\bar\eta)}
\right|\sqrt{g}\propto A^3 \lambda^{11}
\left(1+\lambda^2+O(\lambda^4)\right)\,,
\een
and the Jarlskog invariant $J$ is
\ben
J=\frac{A^2 \bar\eta \lambda^6 (1-A^2\lambda^4) \left(1-\lambda^2 - 2A^2\bar\rho\lambda^4 - A^2 (\bar\eta^2 + (\bar\rho-2) \bar\rho)\lambda^6  + A^4 (\bar\eta^2 + \bar\rho^2)\lambda^8 \right)}{(1-\lambda^2)\left(1 - 2 A^2 \bar\rho \lambda^4 +  A^4 (\bar\eta^2 + \bar\rho^2)\lambda^8\right)^2}=A^2 \bar\eta \lambda^6+O(\lambda^{10})\,.
\een
Inverting this expression to leading order gives the probability distribution for $J$
\ben
p(J)\propto \frac{J}{A\bar\eta^2}\left(1+\left(\frac{J}{A^2\bar\eta}\right)^{1/3}+O(J^{2/3})\right)\,,
\een
which is incompatible with the numerical results. Trying to improve this approximate result by letting $A,\bar\rho$ and $\bar\eta$ take all possible values leads to inconsistencies since the expansion in powers of $J$ contains poles of arbitrary order in $A$. From our present viewpoint, where no mechanism for fixing these parameters close to one is known, the Wolfenstein parametrization seems rather misleading when discussing geometric probability.

\section{Quark Mass Matrices and Gaussian Weighting Functions}
\label{gaussian}

In the previous sections we have focussed on $U(1)^2 \backslash SU(3) / U(1)^2$, the space of Kobayashi-Maskawa matrices, as the space of CP violating parameters. Since $SU(3)$ is compact, this space has finite volume for a natural measure. But the Kobayashi-Maskawa matrix is derived from the Hermitian quark mass matrices, which could be viewed as more fundamental and more directly determined by physics beyond the standard model. In this section, we try to obtain statistics of the Jarlskog invariant $J$ from a random distribution on the space of $3\times 3$ Hermitian matrices. 

\subsection{Distributions on Hermitian matrices}

We follow Sec. \ref{cpviol} and write the quark mass matrices as
\be
UmU^\dagger = {\rm diag} (m_u,m_c,m_t)\,,\qquad
U^\prime m^\prime {U ^\prime}^\dagger = {\rm diag} (m_d,m_s,m_b)\,.
\ee
where $U$ and $U'$ should be thought of as elements of $U(1)^2\backslash SU(3)$. Following \cite{Jarlskog:19851} we normalize the mass matrices by dividing by mass scales $\Lambda$ and $\Lambda'$ (often taken to be the top and bottom quark mass, respectively) which may be chosen for convenience:
\ben
M=U^{\dagger}DU\,,\quad M'={U'}^{\dagger}D'U'\,.
\label{normalised}
\een
The matrices $D$ and $D'$ are now dimensionless quantities, and it is clear that $U$ and $U'$ are only defined up to left multiplication by elements of $U(1)^2$. We consider $\Lambda$ and $\Lambda'$ as arbitrary mass scales, and so we will allow arbitrary eigenvalues for both matrices, instead of fixing one of them to be one.

A natural measure on the space of Hermitian matrices is induced by the metric
\ben
ds^2=\Tr(dM\cdot dM)=\Tr\left(dD\cdot dD\right)+2\Tr\left(\left(dU\,U^{\dagger}\,D\right)^2-\left(dU\,U^{\dagger}\right)^2 D^2 \right)
\een
which is invariant under conjugation under $U(3)$. If we define right-invariant one-forms $\tau_a$ by
\ben
dU\,U^{\dagger}=\im\lambda_a \tau_a\,,
\een
this becomes\footnote{Compare with the 
corresponding result for real matrices given in \cite{Giulini}}
 [with $D\equiv{\rm diag}(D_1,D_2,D_3)$]
\bea
ds^2 & = & \Tr\left(dD\cdot dD\right)-2\tau_a\tau_b\Tr
\left(\lambda_a[D,\lambda_b]D\right)\nn
\\ & = & dD_1^2+dD_2^2+dD_3^2+2\left\{(D_1-D_2)^2(\tau_1^2+\tau_2^2)+(D_1-D_3)^2(\tau_4^2+\tau_5^2)\right.\nn
\\ & & \left.+(D_2-D_3)^2(\tau_6^2+\tau_7^2)\right\}\,.
\eea
The corresponding volume form is
\ben
(D_1-D_2)^2 (D_1-D_3)^2 (D_2-D_3)^2\,dD_1\wedge dD_2\wedge dD_3\wedge \tau_1\wedge \tau_2\wedge \tau_4\wedge \tau_5\wedge \tau_6\wedge \tau_7\,.
\een
As explained above, the measure on the coset $U(1)^2\backslash SU(3)$ is unique and equal to the measure induced from the bi-invariant metric on $SU(3)$. We obtain a Riemannian measure
\ben
DM:=(D_1-D_2)^2 (D_1-D_3)^2 (D_2-D_3)^2 \sin 2x\,\cos^3 y\,\sin y\,\sin 2z\,dD_1\, dD_2\, dD_3\,dx\,dy\,dz\,dw\,dr\,dt
\label{hermit}
\een
on the space of Hermitian $3\times 3$ matrices. The coordinates $(x,y,z,w,r,t)$ on $U(1)^2\backslash SU(3)$ were introduced in Sec. \ref{Su3met}, and we allow arbitrary eigenvalues. (In a fermionic mass term, the sign of the mass has no physical significance, since it can be reversed by multiplying the spinor fields by $\gamma^5$; only $m^2$ enters in physical quantities.) 

From the expressions (\ref{normalised}), it is apparent that each Hermitian matrix with three distinct eigenvalues is associated with six different elements of $\R^3 \times U(1)^2\backslash SU(3)$, related by the action of the discrete group $\frak{S}_3$:
\ben
M=U^{\dagger}DU=(U^{\dagger}P^{-1})PDP^{-1}(PU)=:\tilde{U}^{\dagger}\tilde{D}\tilde{U}\,, \quad P\in \frak{S}_3\,,
\een
where $\frak{S}_3$ is the symmetric group of degree 3 (the dihedral group of order 6, sometimes denoted by $D_3$ or $D_6$) which permutes the canonical basis vectors of $\R^3$. The set of matrices with coinciding eigenvalues has zero measure and hence can be ignored in the present discussion.

Thus we need to consider the space $\R^3 \times (U(1)^2 \times \frak{S}_3)\backslash SU(3)$ instead, restricting the coordinates on the flag manifold to an appropriate range to pick one of the six matrices related by the $\frak{S}_3$ action. We can use the fact that the $\frak{S}_3$ action permutes the rows of an $SU(3)$ matrix to demand that the elements of the third column (see Sec. \ref{Su3met}) satisfy the relation
\ben
|\sin y| \le |\sin x \cos y| \le |\cos x \cos y|\,,
\een
which restricts the coordinates $x$ and $y$ to
\ben
0\le y \le \arctan (\sin x)\,,\quad 0\le x\le \frac{\pi}{4}\,.
\een

Using the natural measure on the flag manifold, we see that this region has precisely one-sixth of the total volume of the flag manifold:
\ben
\frac{\int\limits_0^{\pi/2} dz \int\limits_0^{\pi/4} dx \int\limits_0^{\arctan(\sin x)} dy\,\sin 2x\,\cos^3 y\,\sin y\,\sin 2z}{\int\limits_0^{\pi/2} dz \int\limits_0^{\pi/2} dx \int\limits_0^{\pi/2} dy\,\sin 2x\,\cos^3 y\,\sin y\,\sin 2z}=\frac{1}{6}\,.
\een

 An integral over $\R^3$ with the given measure diverges. We could introduce a cutoff for the quark masses, but then any expectation values for quark masses would strongly contradict observation, as there is no way to explain the observed mass hierarchy.

We therefore choose to introduce a weighting function in the measure which decays sufficiently fast for large positive or negative eigenvalues and is able to reproduce the known hierarchy. The simplest assumption is to take a weighting function of the form
\ben
f\left(\Tr(M^2 A)\right)\,f\left(\Tr((M')^2 A')\right)\,,
\label{eq3}
\een
where $A$ and $A'$ are Hermitian and positive definite, and we shall further assume $[A,A']=0$. By a redefinition of $M$ and $M'$ by unitary conjugation by the same unitary matrix, which leaves $J$ invariant, one can simultaneously diagonalize $A$ and $A'$. For simplicity and ease of technical calculations, we shall choose the function $f$ in (\ref{eq3}) to be a decaying exponential so $M$ and $M'$ are governed by Gaussian distributions.  Our proposal is to fit the diagonal matrices $A$ and $A'$ to the observed quark masses and use the resulting probability distribution for statistics of $J$. 

An integral of a quantity such as $J^2$ becomes\footnote{All odd powers of $J$ again have expectation value zero.}
\bea
\langle J^2 \rangle & = & N \int DM \, DM'\, e^{-\Tr(M^2 A)-\Tr((M')^2 A')} \,J^2(M,M')
\label{jintegral}
\\ & = & N \int_{\R^6} dD \, dD' \int_{((U(1)^2\times \frak{S}_3)\backslash SU(3))^2} DU \, DU'\, e^{-\Tr(D^2 UAU^{\dagger})-\Tr((D')^2 U'A'{U'}^{\dagger})}\,J^2(U,U')\,.\nn
\eea

Here $DU$ and $DU'$ are the measures on $(U(1)^2\times \frak{S}_3)\backslash SU(3)$ and $dD:=(D_1-D_2)^2 (D_1-D_3)^2 (D_2-D_3)^2 \,dD_1\,dD_2\, dD_3$ etc., and the normalization factor $N$ is defined by
\ben
\frac{1}{N}:=\int_{\R^6} dD \,dD' \int_{((U(1)^2\times\frak{S}_3)\backslash SU(3))^2} DU \, DU'\, e^{-\Tr(D^2UAU^{\dagger})-\Tr((D')^2U'A'{U'}^{\dagger})}\,.
\een
From $J=\Im \bigl (V_{11} \,V_{22}\, V_{12}^* \, V_{21}^* \bigr )$ and $V=U{U'}^{\dagger}$, we have
\ben
J(U,U')=\sum_{a,b,c,d=1}^3\Im\bigl( U_{1a}U_{2b}U_{1c}^*U_{2d}^*{U'}^*_{1a}{U'}^*_{2b}{U'}_{2c}{U'}_{1d}\bigr)\,.
\een

At this point, it is perhaps instructive to note that setting $A$ equal to the identity would split the integral (\ref{jintegral}) into a product of an integral over the eigenvalues which just gives a constant and an integral of $J^2$ over $((U(1)^2\times\frak{S}_3)\backslash SU(3))^2$. Since all even powers of $J$ are invariant under the $\frak{S}_3$ action on $U$ and $U'$, this can be replaced by an integral over $(U(1)^2\backslash SU(3))^2$ if averages are concerned. By the arguments presented in Sec. \ref{flagman}, a change of coordinates reduces this to a single integration over a flag manifold, and one recovers the results of Sec. \ref{flagman} for expectation values of powers of $J$.

The introduction of more general diagonal matrices $A$ and $A'$ means that the invariance of the measure $DM\,DM'$ under separate conjugation of $M$ and $M'$ by arbitrary elements of $U(3)$, \ie under the action of $U(3)\times U(3)$, is broken down to the action of the diagonal subgroup $U(1)^2 \times U(1)^2$ which commutes with $A$ and $A'$. We find that this symmetry breaking is necessary to obtain a distribution that reproduces different expectation values for squared quark masses.

It should be clear from (\ref{jintegral}) that multiplying $A$ (or $A'$) by a constant is the same as rescaling the eigenvalues $D_i$ (or $D_i'$) and so amounts to a rescaling of $\Lambda$ (or $\Lambda'$). We can therefore, without any loss of generality, choose
\ben
A=\left(\begin{matrix}
1 & 0 & 0 \\ 0 & 1/\mu_c^2 & 0 \\ 0 & 0 & 1/\mu_u^2
\end{matrix}\right),\quad
A'=\left(\begin{matrix}
1 & 0 & 0 \\ 0 & 1/\mu_s^2 & 0 \\ 0 & 0 & 1/\mu_d^2
\end{matrix}\right)\, ,
\label{amatrices}
\een
where $\mu_c, \mu_u, \mu_s$, and $\mu_d$ are dimensionless parameters that we are free to choose so as to reproduce the observed quark masses as expectation values. (In the case of an exponential $\exp(-\Tr(D^2 A))$, these would of course be equal to the respective quark masses, expressed in units where $\Lambda=m_t$ and $\Lambda'=m_b$.) Because of experimental uncertainties in the up and quark masses, one can modify this distribution to reproduce different values for these masses.

It seems practically impossible to evaluate the integral (\ref{jintegral}), as the expression for $J$ in terms of coordinates on $((U(1)^2\times \frak{S}_3)\backslash SU(3))^2$ is too complicated to be given explicitly. However, since
\ben
\Tr(D^2 U A U^{\dagger})=\sum_a D_a^2 \sum_c A_c |U_{ac}|^2=: \sum_a D_a^2 \xi_a\,, \quad \Tr((D')^2 U' A' {U'}^{\dagger})=: \sum_a (D'_a)^2 \xi'_a
\een
with
\ben
\xi_1= A_1 \cos^2 y \cos^2 z + A_2 \cos^2 y \sin^2 z + A_3 \sin^2 y\,, \quad \xi'_1= A'_1 \cos^2 y' \cos^2 z' + A'_2 \cos^2 y' \sin^2 z' + A'_3 \sin^2 y'\,,
\een
and we assume $A_3\gg 1$ and $A_3'\gg 1$, the integrand is negligibly small unless $y\approx 0$ and $y'\approx 0$. We use this to approximate the integrals over $y$ and $y'$:
\bea
& & \int\limits_{0}^{\arctan(\sin x)} dy \int\limits_{0}^{\arctan(\sin x')} dy' \, \cos^3 y \,\sin y\, \cos^3 y' \,\sin y'\, e^{-\Tr(D^2 UAU^{\dagger})-\Tr((D')^2 U'A'{U'}^{\dagger})}\,J^2(U,U') \nn
\\ & \approx & \int\limits_{0}^{\arctan(\sin x)} dy \int\limits_{0}^{\arctan(\sin x')} dy' \, y\, y'\, e^{-A_3 y^2 -A_3' (y')^2}\left(e^{-\Tr(D^2 UAU^{\dagger})-\Tr((D')^2 U'A'{U'}^{\dagger})}\,J^2(U,U')\right)\big|_{y=y'=0} \nn
\\ & \approx & \frac{1}{4A_3 A_3'}\left(e^{-\Tr(D^2 UAU^{\dagger})-\Tr((D')^2 U'A'{U'}^{\dagger})}\,J^2(U,U')\right)\big|_{y=y'=0}\,.
\eea

It turns out that this is independent of $w$ and $w'$. Constant prefactors such as $1/4A_3 A'_3$ appearing in both numerator and denominator can be dropped, and so we have
\ben
\langle J^2 \rangle \approx \frac{\int_{\R^6} dD \,dD' \int d^4 x\int d^4 x'\,\sin 2x \sin 2z \sin 2x' \sin 2z' \left(e^{-\Tr(D^2 UAU^{\dagger})-\Tr((D')^2 U'A'{U'}^{\dagger})}\,J^2(U,U')\right)\big|_{y=y'=0}}{\int_{\R^6} dD \, dD' \int d^4 x\int d^4 x'\,\sin 2x \sin 2z \sin 2x' \sin 2z' \left(e^{-\Tr(D^2 UAU^{\dagger})-\Tr((D')^2 U'A'{U'}^{\dagger})}\right)\big|_{y=y'=0}}\,,
\label{newjint}
\een
where
\ben
\int d^4 x\equiv \int\limits_0^{\pi/4}dx\int\limits_0^{\pi/2}dz \int\limits_0^{2\pi}dr\int\limits_0^{2\pi}dt
\een
and similarly for $\int d^4 x'$.

Now we can integrate over both copies of $\R^3$ in (\ref{newjint}), using
\bea
& &\int\limits_{-\infty}^{\infty}dD_1\int\limits_{-\infty}^{\infty}dD_2\int\limits_{-\infty}^{\infty}dD_3\,(D_1-D_2)^2(D_1-D_3)^2(D_2-D_3)^2 e^{-\xi_1 D_1^2-\xi_2D_2^2-\xi_3D_3^2}\nn
\\& = &\frac{3\pi^{3/2}}{8 \xi_1^{5/2}\xi_2^{5/2}\xi_3^{5/2}}\left(\xi_1^2(\xi_2+\xi_3)+\xi_2^2(\xi_1+\xi_3)+\xi_3^2(\xi_2+\xi_1)-2\xi_1\xi_2\xi_3\right)\,.
\eea

The explicit expression for $J$ at $y=y'=0$ is
\bea
J(U,U')\big|_{y=y'=0} & = &\frac{1}{4}s_{2x} s_{2x'} \left\{c^2_{z'} s^3_{z} s_{z'}\sin(3 \hat{r} + \hat{t})  + c^3_{z} c_{z'} s^2_{z'}\sin(3 \hat{r} - \hat{t}) \right.\nn
\\& &  - c^2_{z} s_{z} s_{z'} (c^2_{z'} \left[\sin(3 \hat{r} + \hat{t}) + \sin(3\hat{r} -3\hat{t})\right] - s^2_{z'}\sin(3\hat{r} + \hat{t}) )\nn
\\& & \left. + c_{z} c_{z'} s^2_{z} (c^2_{z'} \sin(3\hat{r} - \hat{t}) -  s^2_{z'}\left[\sin(3 \hat{r} + 3\hat{t}) + \sin(3 \hat{r} - \hat{t})\right])\right\}
\label{jexp}
\eea
where $s_x=\sin x, c_{z'}=\cos z'$, etc., $\hat r= r-r'$, and $\hat t=t-t'$. Integrating (\ref{jexp}) over $r,r',t$, and $t'$ indeed gives zero, which is why we choose to use $J^2$.

\subsection{Results and dependence on quark masses}

We need to determine the parameters appearing in the matrices $A$ and $A'$ in (\ref{amatrices}). We first observe that expectation values for squared mass matrices take the relatively simple form
\ben
\langle D_1^2 \rangle \approx \frac{\int_{\R^3} dD\, D_1^2 \int\limits_0^{\pi/4} dx \int\limits_0^{\pi/2} dz\,\sin 2x \sin 2z  \left(e^{-\Tr(D^2 UAU^{\dagger})}\right)\big|_{y=0}}{\int_{\R^3} dD\,  \int\limits_0^{\pi/4} dx \int\limits_0^{\pi/2} dz\,\sin 2x \sin 2z  \left(e^{-\Tr(D^2 UAU^{\dagger})}\right)\big|_{y=0}}\,.
\een
The denominator is explicitly
\ben
I_D:=\int\limits_0^{\pi/4} dx \int\limits_0^{\pi/2} dz\,\sin 2x \sin 2z \,\frac{3\pi^{3/2}}{8 \xi_1^{5/2}\xi_2^{5/2}\xi_3^{5/2}}\left(\xi_1^2(\xi_2+\xi_3)+\xi_2^2(\xi_1+\xi_3)+\xi_3^2(\xi_2+\xi_1)-2\xi_1\xi_2\xi_3\right)\,,
\label{denom}
\een
where
\bea
& \xi_1=A_1 \cos^2 z + A_2 \sin^2 z\,,\quad \xi_2=A_1 \cos^2 x \sin^2 z + A_2 \cos^2 x \cos^2 z + A_3 \sin^2 x\,,\nn
\\& \xi_3=A_1 \sin^2 x \sin^2 z + A_2 \sin^2 x \cos^2 z + A_3 \cos^2 x\,,
\eea
with $A_3 \gg A_2 \gg A_1$. We notice that all $\xi_a$ are nonzero for all values of $x$ and $z$. Furthermore, the integral is dominated by very small $x$ and $z$ (we cannot have $x=\frac{\pi}{2}$), and we can approximate $I_D$ well by only keeping the terms of leading order in $x$ and $z$ in the trigonometric functions, and 
\ben
\xi_1^2(\xi_2+\xi_3)+\xi_2^2(\xi_1+\xi_3)+\xi_3^2(\xi_2+\xi_1)-2\xi_1\xi_2\xi_3\approx A_3^3 x^2 + A_3^2 A_2\,,
\een
which are the leading terms (as we shall see, the first of these is effectively also of order $A_3^2 A_2$):
\bea
I_D & \approx &\frac{3\pi^{3/2}}{8}\int\limits_0^{\pi/4} dx \int\limits_0^{\pi/2} dz\,4 x z \,(A_3^3 x^2 + A_3^2 A_2)(A_1+A_2 z^2)^{-5/2}(A_2+A_3 x^2)^{-5/2}A_3^{-5/2}\nn
\\ & \approx &\frac{3\pi^{3/2}}{8}\int\limits_0^{\infty} dX \int\limits_0^{\infty} dZ\,(A_3^3 X + A_3^2 A_2)(A_1+A_2 Z)^{-5/2}(A_2+A_3 X)^{-5/2}A_3^{-5/2}\nn
\\ & = &\frac{3\pi^{3/2}}{8}\int\limits_0^{\infty} dX (A_3^3 X + A_3^2 A_2)(A_2+A_3 X)^{-5/2}A_3^{-5/2}\cdot\frac{2}{3A_1^{3/2}A_2}\nn
\\ & = &\frac{3\pi^{3/2}}{8}\frac{2}{3A_1^{3/2}A_2}\left(\frac{2}{3} A_2^{-1/2}A_3^{-3/2}+\int\limits_0^{\infty} dX \,\frac{2}{3A_3^{1/2}}(A_2+A_3 X)^{-3/2}\right)\nn
\\ & = & \frac{\pi^{3/2}}{4 A_1^{3/2}A_2}\left(\frac{2}{3} A_2^{-1/2}A_3^{-3/2}+\frac{4}{3A_3^{3/2}A_2^{1/2}}\right) = \frac{\pi^{3/2}}{2 A_1^{3/2}A_2^{3/2}A_3^{3/2}}\,.
\eea
Similarly, we find
\bea
I_D\langle D_1^2 \rangle & \approx & \frac{15\pi^{3/2}}{16}\int\limits_0^{\infty} dX \int\limits_0^{\infty} dZ\,(A_3^3 X + A_3^2 A_2)(A_1+A_2 Z)^{-7/2}(A_2+A_3 X)^{-5/2}A_3^{-5/2}\nn
\\ & = & \frac{15\pi^{3/2}}{16}\int\limits_0^{\infty} dX \,(A_3^3 X + A_3^2 A_2)(A_2+A_3 X)^{-5/2}A_3^{-5/2}\cdot\frac{2}{5A_2 A_1^{5/2}} \nn
\\ & = & \frac{3\pi^{3/2}}{4 A_1^{5/2}A_2^{3/2}A_3^{3/2}}\,,
\eea
hence
\ben
\langle D_1^2 \rangle \approx \frac{3}{2A_1}\,.
\een
Redoing the same calculation for $D_2$ and $D_3$ gives
\ben
\langle D_2^2 \rangle \approx \frac{1}{2A_2}\,,\quad \langle D_3^2 \rangle \approx \frac{1}{2A_3}\,.
\een
There is a relative factor of 3 which has to be taken into account when determining $A$ and $A'$.

Because of the dependence of masses on the energy scale in quantum field theory, described by the renormalization group, there is some ambiguity in what is meant by the ``quark masses" we want to reproduce. Following \cite{rosner}, for example, we take all the quark masses evolved to the scale of the $Z$ boson mass. These are given in \cite{massref}:
\bea
(m_u,m_c,m_t)=(1.27_{-0.42}^{+0.50}\;{\rm MeV},\; 0.619 \pm 0.084\;{\rm GeV},\;171.7 \pm 3.0\;{\rm GeV})\,; \nn
\\ (m_d,m_s,m_b)=(2.90_{-1.19}^{+1.24}\;{\rm MeV},\; 55_{-15}^{+16}\;{\rm MeV},\;2.89 \pm 0.09\;{\rm GeV})\,.
\label{qmass}
\eea
We use the central values
\ben
(m_u,m_c,m_t):=(1.27\;{\rm MeV},\; 0.619\;{\rm GeV},\;171.7\;{\rm GeV})\,; (m_d,m_s,m_b):=(2.9\;{\rm MeV},\; 55\;{\rm MeV},\;2.89\;{\rm GeV})\,.
\label{cvalues}
\een

The mass scales $\Lambda$ and $\Lambda'$ are fixed by setting $\langle D_1^2 \rangle=(m_t/\Lambda)^2$ and $\langle (D'_1)^2 \rangle=(m_b/\Lambda')^2$. By comparing the results obtained by numerical integration with the values we want to reproduce, we can then fix the parameters $\mu_c,\mu_u,\mu_s$ and $\mu_d$.

In the case of the positively charged top, charm and up quarks, which exhibit a more extreme quark mass hierarchy, we find that numerical calculations (using {\it Mathematica}) reproduce the results we have obtained analytically very well (see Table \ref{results}). For the negatively charged quarks, we find numerically that we have to use relative factors different from 3 to reproduce the observed masses. Comparing the numerical results with (\ref{cvalues}), we fix the parameters appearing in $A$ and $A'$ to
\bea
\mu_c^2 = 3\left(\frac{m_c}{m_t}\right)^2\approx 3.90\times 10^{-5}\,,\quad \mu_u^2 = 3\left(\frac{m_u}{m_t}\right)^2\approx 1.64\times 10^{-10}\,,\nn
\\ \mu_s^2 = \frac{3}{2}\left(\frac{m_s}{m_b}\right)^2\approx 5.43\times 10^{-4}\,,\quad \mu_d^2 = \frac{12}{5}\left(\frac{m_d}{m_b}\right)^2\approx 2.42\times 10^{-6}\,.
\label{muvalues}
\eea

As a brief side remark, we see that the dominant contributions to these integrals come from the regions
\ben
y\approx\sqrt{\frac{1}{A_3}}\,,\quad y'\approx\sqrt{\frac{1}{A'_3}}\,,\quad x\approx\sqrt{\frac{A_2}{A_3}}\,,\quad z\approx\sqrt{\frac{A_1}{A_2}}\,,\quad x'\approx\sqrt{\frac{A'_2}{A'_3}}\,,\quad z'\approx\sqrt{\frac{A'_1}{A'_2}}\,,
\label{regions}
\een
and these values are all small compared to one. We can therefore give rough estimates for magnitudes of individual elements of the Kobayashi-Maskawa matrix.

In the standard convention the ordering of the quark families is $(u,c,t)$ and not $(t,c,u)$ as used in (\ref{amatrices}), which means that in our parametrization,
\ben
|(U{U'}^{\dagger})_{13}|=|V_{td}|\,,\quad |(U{U'}^{\dagger})_{12}|=|V_{ts}|\,,\quad |(U{U'}^{\dagger})_{23}|=|V_{cd}|\,.
\een
Since all of the numbers in (\ref{regions}) are small, we only keep leading terms in the angles on $U$ and $U'$:
\bea
|(U{U'}^{\dagger})_{13}|\approx|x'(z'-z)-e^{iw'}y'+\ldots|\approx x'z'\approx \mu_d \approx 2 \times 10^{-3}\,, \nn
\\ |(U{U'}^{\dagger})_{12}|\approx z'\approx \mu_s\approx 0.02 \,,\quad |(U{U'}^{\dagger})_{23}|\approx x'\approx \frac{\mu_d}{\mu_s}\approx 0.07\,. 
\eea
Experimental values are \cite{particledb}
\ben
|V_{td}|=(8.14_{-0.64}^{+0.32}) \times 10^{-3}\,,\quad |V_{ts}|=(41.61_{-0.78}^{+0.12}) \times 10^{-3}\,,\quad |V_{cd}|=0.2271_{-0.0010}^{+0.0010}\,.
\een
Our rough estimates reproduce the right ordering of the three parameters and are accurate to factors of order a few. A more careful analysis would involve computing expectation values for these parameters in the distribution we have assumed.

We return to the task of computing the expectation value of $J^2$. In order to obtain an analytical expression, we use the fact that the main contribution to the integral (\ref{newjint}) will come from small $z$ to only take the term in (\ref{jexp}) that is nonzero at $z=0$. Averaging over $r,t,r'$,and $t'$ gives a factor of 1/2, as one might have expected, and therefore we use
\ben
J^2_{{\rm small}\;z}:=\frac{1}{2}\sin^2 x\cos^2 x\sin^2 x'\cos^2 x'\cos^2 z'\sin^4 z'
\een
for our calculations. The numerator of (\ref{newjint}) is the product (using again that only small $z$ contributes)
\bea
& &\frac{9\pi^3}{32}\times \int\limits_0^{\pi/2} dz \, \frac{2z}{(A_1 + A_2 z^2)^{5/2}} \times \int\limits_0^{\pi/2} dz' \, \frac{\sin 2z' \cos^2 z'\sin^4 z'}{(A_1 \cos^2 z'+ A_2 \sin^2 z')^{5/2}} \nn
\\ & \times &\int\limits_0^{\pi/4} dx \,\frac{\sin 2x\,\sin^2 x\,\cos^2 x(A_3^3 \cos^2 x \sin^2 x + A_3^2 A_2(\cos^6 x + 2 \cos^2 x \sin^2 x + \sin^6 x))}{(A_2 \cos^2 x + A_3 \sin^2 x)^{5/2} (A_3 \cos^2 x + A_2 \sin^2 x)^{5/2}}
\\ & \times &\int\limits_0^{\pi/4} dx' \,\frac{\sin 2x'\,\sin^2 x'\,\cos^2 x'((A'_3)^3 \cos^2 x' \sin^2 x' + (A'_3)^2 A'_2(\cos^6 x' + 2 \cos^2 x' \sin^2 x' + \sin^6 x'))}{(A'_2 \cos^2 x' + A'_3 \sin^2 x')^{5/2} (A'_3 \cos^2 x' + A'_2 \sin^2 x')^{5/2}} \nn
\eea
The first two factors are $2/(3A_1^{3/2}A_2)$ and $4/(3\sqrt{A_2'}(\sqrt{A_1'}+\sqrt{A_2'})^4)$, respectively; for the other two (which are identical) we change variables to $X=\cos^2 x$ to obtain
\ben
\int\limits_{1/2}^{1} dX \,\frac{X(1-X)(A_3^3 X(1-X) + A_3^2 A_2(X^2 - X + 1))}{(A_2 X + A_3 (1-X))^{5/2} (A_3 X + A_2 (1-X))^{5/2}}\approx \frac{1}{A_3^2}\left(\arctan\left(\frac{1}{2}\sqrt{\frac{A_3}{A_2}}\right)-2\sqrt{\frac{A_2}{A_3}}\right)\,,
\een
where we are dropping corrections of order $\frac{A_2}{A_3}$. Putting everything together, we obtain
\bea
\langle J^2_{{\rm small}\;z}\rangle & \approx &\frac{(A_1')^{3/2}A_2'\sqrt{A_2}}{\sqrt{A_3 A_3'}(\sqrt{A'_1}+\sqrt{A'_2})^4}\left(\arctan\sqrt{\frac{A_3}{4A_2}}-\sqrt{\frac{4A_2}{A_3}}\right)\left(\arctan\sqrt{\frac{A'_3}{4A'_2}}-\sqrt{\frac{4A'_2}{A'_3}}\right)
\label{japprox}
\\ & = & \frac{\frac{4}{\sqrt{15}}m_u\,m_d\,m_b}{m_c \,m_s^2\left(1+\sqrt{\frac{2}{3}}\frac{m_b}{m_s}\right)^4}\left(\arctan\frac{m_c}{2m_u}-\frac{2m_u}{m_c}\right)\left(\arctan\sqrt{\frac{5}{32}}\frac{m_s}{m_d}-\sqrt{\frac{32}{5}}\frac{m_d}{m_s}\right)\,,\nn
\eea
where the numerical factors appearing in the last line come from the different factors chosen in (\ref{muvalues}). Note that the top quark mass does not appear in this approximate result.

For numerical calculations we use both the simplified expression $J^2_{{\rm small}\;z}$ and the expression for $J$ given in (\ref{jexp}). We find that for the first quantity, the numerically evaluated expectation value $\langle J^2_{{\rm small}\;z}\rangle$  is about 7/6 of (\ref{japprox}), and the numerical result for $\langle J^2 \rangle$ (taken at $y=y'=0$) is
\ben
\langle J^2 \rangle \approx 5.28\times 10^{-9}\,,
\een
which gives
\ben
\Delta J = \sqrt{\langle J^2 \rangle} \approx 7.27 \times 10^{-5}
\een
which is much closer to the observed value than any of the previously obtained results. Assuming a Gaussian distribution for $J$ which is peaked at zero, the probability of finding a small $J$, in the sense of Sec. \ref{fine}, is now
\ben
P_{{\rm mass}}(|J|\le 10^{-4})\approx 83\%\,,
\een 
whereas the probability of finding a $J$ which is even smaller than the observed value is
\ben
P_{{\rm mass}}(|J|\le 3\times 10^{-5})\approx 32\%\,.
\een 
The observed value for $J$ can no longer be viewed as being finely tuned if the distribution used in our calculations is assumed.

\begin{table}[htp]
\caption{{\small Analytical and numerical results for integrals of interest.}}
\begin{center}
\begin{tabular}{c | c | c | c | c }
 Quantity & $I_D$ (over $x, z$) & $I_D\langle D_1^2\rangle$  & $I_D\langle D_2^2\rangle$  & $I_D\langle D_3^2\rangle$ 
\\\hline
 Analytical result & $ 1.43 \times 10^{-21}$ & $ 2.14 \times 10^{-21}$  & $ 2.78 \times 10^{-26}$  & $ 1.17 \times 10^{-31}$ 
\\\hline
 Numerical result & $ 1.43 \times 10^{-21}$ & $ 2.14 \times 10^{-21}$  & $ 2.78 \times 10^{-26}$  & $ 1.18 \times 10^{-31}$ 
\\\hline\hline
 Quantity & $I'_D$ (over $x', z'$) & $I'_D\langle (D'_1)^2\rangle$  & $I'_D\langle (D'_2)^2\rangle$  & $I'_D\langle (D'_3)^2\rangle$ 
\\\hline
 Analytical result & $ 1.32 \times 10^{-13}$ & $ 1.99 \times 10^{-13}$  & $ 3.60 \times 10^{-17}$  & $ 1.60 \times 10^{-19}$ 
\\\hline
 Numerical result & $ 1.32 \times 10^{-13}$ & $ 1.98 \times 10^{-13}$  & $ 7.28 \times 10^{-17}$  & $ 1.99 \times 10^{-19}$ 
\\\hline\hline
 Quantity & $\tilde{I}_D$ (over $x,z,x', z'$) & $\langle J^2_{{\rm small}\;z}\rangle$  & $\langle J^2\rangle$
\\\hline
 Analytical result & $ 1.89 \times 10^{-34}$ & $ 3.22 \times 10^{-9}$  & ---
\\\hline
 Numerical result & $ 1.88 \times 10^{-34}$ & $ 3.73 \times 10^{-9}$  & $ 5.28 \times 10^{-9}$ 
\\\hline\hline
\end{tabular}
\end{center}
\label{results}
\end{table}

To test the sensitivity of our results to changes in the parameters, we take values at the upper or lower limit in (\ref{qmass}) and try to find the highest and lowest values for $\langle J^2\rangle$. We find that setting 
\ben
(m_u,m_c,m_t):=(0.85\;{\rm MeV},\; 0.535\;{\rm GeV},\;174.7\;{\rm GeV})\,; (m_d,m_s,m_b):=(1.71\;{\rm MeV},\; 40\;{\rm MeV},\;2.98\;{\rm GeV})
\een
gives
\ben
\langle J^2 \rangle \approx 1.86 \times 10^{-9}
\een
and
\ben
\Delta J = \sqrt{\langle J^2 \rangle} \approx 4.31 \times 10^{-5}\,,
\een
whereas setting
\ben
(m_u,m_c,m_t):=(1.77\;{\rm MeV},\; 0.535\;{\rm GeV},\;168.7\;{\rm GeV})\,; (m_d,m_s,m_b):=(4.14\;{\rm MeV},\; 71\;{\rm MeV},\;2.8\;{\rm GeV})
\een
gives
\ben
\langle J^2 \rangle \approx 1.52 \times 10^{-8}
\een
and
\ben
\Delta J = \sqrt{\langle J^2 \rangle} \approx 1.23 \times 10^{-4}\,.
\een
Since the former choice makes $J$ appear more typical, these results are perhaps an indication that the correct values for the (up, down, and strange) quark masses are probably closer to the lower than to the upper bounds given in (\ref{qmass}). Also, even the greatest possible value for $\Delta J$ is significantly lower than any of the values obtained in previous sections.

In this section, we have established that assuming the observed hierarchy in quark masses in a Gaussian distribution over the space of mass  matrices gives expectation values for $J^2$ which are small enough to regard the observed value as ``natural'' and not finely tuned. This statistical observation seems to open up the possibility that the same mechanism that is responsible for the apparently unlikely hierarchy in quark masses might also explain why the observed value for $|J|$ is so small.

A more detailed analysis including a probability density for $|J|$ for this distribution is left to future work, since the numerical methods used here do not give sufficiently accurate results.
  
\section{Extension to Neutrinos}
\label{neutrinos}

In this section we review the case of neutrino masses, highlighting the difference between Majorana and Dirac masses and commenting on some recently made suggestions in the literature that one could distinguish the two cases by gravitational effects. In contrast to the physical situation which is at present rather unclear, the mathematical problem of obtaining a measure on the space of mixing matrices is in this case simpler, since one considers the flag manifold $U(1)^2\backslash SU(3)$. 

\subsection{Neutrino masses}

The spectrum of neutrinos and their masses and their nature, Majorana
or Dirac,  is currently not
well known. 
Since not all of the material may be familiar
to all readers we shall review 
some basic facts about Dirac and Majorana masses in a framework
which is  sufficiently general to encompass all  likely possibilities.   
We  find it helpful to use a Majorana notation, 
but no loss of generality  thereby
results since, if one starts with complex Weyl notation,
one may always take real and imaginary parts. Alternatively,
given a treatment  in terms of Majorana spinors, one may always 
transcribe it into Weyl notation.  In order to simplify the analysis
we depart from common practice in phenomenological particle physics and 
adopt the spacetime signature $(-+++)$ for all spinors. This has the advantage
that   all  gamma
matrices may  be taken to be  real, $^t$ denotes 
transpose.  $C=-C^t$ is the charge conjugation matrix and 
$\gamma^5 =-(\gamma ^5)^t = \gamma ^0 \gamma ^1 \gamma ^2 \gamma ^3$
so that $(\gamma^5)^2 =-1$. If required,   
a concrete representation
is given by
\bea
\gamma^0 &=&  \begin{pmatrix}
 0 & 1 & 0 & 0 \cr  -1 & 0 & 0 & 0 \cr
  0 & 0 & 0 & -1 \cr  0 & 0 & 1 & 0 \cr 
\end{pmatrix}\,,\qquad
\gamma^1  =\begin{pmatrix}
 0 & 1 & 0 & 0 \cr  1 & 0 & 0 & 0 \cr
  0 & 0 & 0 & 1 \cr  0 & 0 & 1 & 0 \cr 
\end{pmatrix}\,,\nn\\
\gamma^2 &=& \begin{pmatrix}
 1 & 0 & 0 & 0 \cr  0 & -1 & 0 & 0 \cr
  0 & 0 & 1 & 0 \cr  0 & 0 & 0 & -1 \cr 
\end{pmatrix}\,,\qquad
\gamma^3  =\begin{pmatrix} 0 & 0 & 0 & 1 \cr  0 & 0 & -1 & 0 \cr
  0 & -1 & 0 & 0 \cr  1  & 0 & 0 & 0 \cr 
\end{pmatrix}\,.
\eea
In this representation we may take $C= \gamma_0$, and it is often useful
to note that $\gamma^0$ is 
antisymmetric while  $\gamma^1,\gamma^2$ and $\gamma ^3$ are
symmetric.

The mass matrix of a set of fermions is defined by imagining
setting to zero that part of the effective or large distance 
Lagrangian containing couplings  to all gauge interactions except gravity.
In fact this is how information about neutrino masses is obtained.
One observes mixing as they pass from the sun to the earth
and upper bounds on their masses have been  obtained by
observing their  arrival times from distant supernova 1987a.

The most general effective Lorentz-invariant Lagrangian
for a system of $k$ free Majorana fermions $\psi ^i$, $i=1,2,\dots ,k$, is 
 \ben
L = \half \psi^t C \Dslash \psi - \half  \psi ^t C \bigl 
(m_1+m_2 \gamma ^5 \bigr ) \psi\,, 
\een
where $\psi$ should be thought of as a $k$-dimensional
column vector all of whose entries are four component real
Majorana spinors, and  $m_1$ and $m_2$ are real symmetric $k\times k$ matrices.
Note that at this stage $k$ may be even or odd.
We have made use of the fact that we may diagonalize 
the  kinetic term using $GL(k,\mathbb{R})$
transformations. We are still allowed    
$SO(k) \subset GL(k,\mathbb{R})$ transformations
\ben
\psi \rightarrow O \psi\,,\qquad O ^t O =1\,,\een
where
\ben
O= \exp\left(\omega_{ij}\right)\,,\qquad \omega_{ij}=-\omega_{ji}\,.
\een

The kinetic term, but not the mass term  is also  invariant under   
chiral rotations
\ben
\psi \rightarrow  P \psi\,,\een 
\ben
P= \exp\left(\nu_{ij} \gamma ^5\right)\,,\qquad \nu_{ij}= \nu_{ji}\,.
\een
Combining these two sets of transformations we see that 
the kinetic term, but not the mass term is in fact invariant
under the action of $U(k)$, \ie under
\ben
\psi \rightarrow  S \psi\,, 
\een
\ben
S= \exp {\bigl(  \omega_{ij} + \nu_{ij} \gamma ^5 \bigr ) }\,.
\een
The $U(k)$ invariance is perhaps more obvious
if one uses a Weyl basis. Since
\ben
\bigl( \gamma ^5 \bigr )^2 =-1\,,
\een 
one may regard  $\gamma ^5$ as providing a complex structure
on the space of $4k$  real dimensional   Majorana spinors, 
converting it to the  $2k$ complex dimensional
space of positive chirality  Weyl spinors for which  
\ben
\gamma^5 = {\rm i}\,.
\een
Clearly $S$  then becomes the   exponential of the $k\times k$ 
anti-Hermitian matrix
\ben
 \omega_{ij} + {\rm i} \nu_{ij}\,.
\een  
Thus
\ben
S  S^\dagger = 1\,.
\een
The mass matrix is then a complex symmetric matrix
\ben
m= m_1+i m_2 \,, 
\een
and under a $U(k)$ transformation
\ben
m \rightarrow S^{\dagger} m S\,.
\een
At this point we invoke the result of Zumino \cite{Zumino}
that $S$ may be chosen to render the matrix $m$ diagonal with
real non-negative  entries $m_i$.

In the general case, all the masses $m_i$ are distinct. They are then
said to be of Majorana type. However, it may happen that two
masses, $m_1$ and $m_2$ say, coincide. One may then
combine $\psi^1, \psi ^2$ into a Dirac spinor. One then has the case
of  a Dirac mass. For quarks and charged leptons all 
masses are of Dirac type. For neutrinos, however, it is
not yet known of  what type they are, nor indeed how many.
A simple assumption is that $k=6$, with  three 
having very heavy masses and three having very light masses.
This corresponds to the so-called \lq \lq seesaw mechanism\rq\rq. 
Of course the very light neutrinos may be combined
into three Weyl neutrinos and are taken to be massless in the standard
model.     

From the analysis above  it follows that in the general case
when all masses $m_i$ are distinct, there is a unique basis for the neutrino
states, determined by their inertial  motion. 
If, however, two or more masses coincide, then 
the basis becomes ambiguous up to rotations
of the components with equal masses. In the case that $k$ is even
and there are $[{k \over 2} ]$ distinct pairs of coincident masses
the basis is arbitrary up to the action of 
$T^{[{k \over 2}] } \equiv U(1)^{[{k \over 2}]  }$.

Any mixing matrix
taking one to a basis  which is preferred from the point
of their nongravitational gauge interactions will be ambiguous
to the extent that the inertial  basis and the gauge basis
are ambiguous. For that reason, in general, a mixing matrix belongs
to a double quotient.

\subsubsection {The  Universality of free fall}
In our discussion above we have referred
to the interactions of neutrinos with gravitational fields.
Of course neutrinos observed to be coming from
the   sun, or the supernova 1987a are traveling
so fast that the effects of gravity on them are negligible.
However, it has been suggested that it is in principle
possible to distinguish  Majorana from  Dirac masses
by their  behavior in the  gravitational
fields of  rotating  objects \cite{Singh1,Singh2,Singh3}. 
Our analysis above shows
that unless there are gauge interactions such as might correspond
to neutrino magnetic or electric dipole moments
this is not so, as long as the coupling to gravity is 
\lq \lq minimal.\rq \rq If so one simply uses for $\nabla$
the standard Levi-Civita covariant derivative acting on spinors.       

Assuming  that the mass matrix $m$ is independent of position,
we may take it to be everywhere  real and diagonal.
Thus each component $ \psi^i$ of the inertial basis propagates
independently.
One may iterate the Dirac equation and use the cyclic Bianchi
identity in a curved space to get (reinstating powers of $\hbar$)
\ben
-\hbar^2 \nabla ^2 \psi ^i + {1\over 4} \hbar^2 R \psi^i +  
m_i^2 \psi ^i =0\,. \label{iterated}
\een
If the effects of curvature are negligible on the scale of the Compton wavelength,
\ben
\frac{\hbar^2}{m_i^2} \ll L_c^2\,,
\een
the second term may be dropped and one obtains the Klein-Gordon equation for each component.

As is well known, there is no ``gyro-magnetic'' coupling between the
spin and the Ricci or Riemann tensors  \cite{Peres}. 
To proceed, one may pass to a 
Liouville-Green-Jeffreys-Wentzel-Kramers-Brillouin (L-G-J-W-K-B)
approximation of the form 
\ben
\psi^i = \chi^i  e^{i S/\hbar}\,.
\een
One obtains from the original Dirac equation
\ben
\bigl( i\gamma ^\mu \partial _\mu S + m_i \bigr) \chi^i =0 
\label{Dirac} 
\een
and 
\ben
\gamma^{\mu} \nabla _\mu \chi ^i =0\,. \label{spinor}
\een
It follows from (\ref{Dirac}) that
\ben
{\rm det} \Bigl ( i \gamma ^\mu  \partial _\mu S  + m_i \Bigr )=0\,. \label{det} 
\een
Evaluation of the determinant in (\ref{det}) gives
the Hamilton-Jacobi equation
\ben
 g^{\mu \nu} \partial _\mu\, S \partial _\nu S + m_i^2   =0\,. 
\label{HJ
}\een
This shows that the orthogonal trajectories defined by
\ben
m_i { d x ^\mu \over d \tau} = g^{\mu \nu} \partial _\mu S  
\een
are timelike geodesics. The same conclusion follows
by applying the L-G-J-W-K-B approximation
to the second order iterated Dirac equation (\ref{iterated}).  

The iterated Dirac equation also gives
\ben
\partial _\mu S \nabla ^\mu \chi ^i = - \frac{1}{2}(\nabla^2 S) \chi^i\,.
\een
The same result may be obtained by differentiating
(\ref{Dirac}) and using (\ref{spinor}).
Thus the spinor $\chi^i$ is parallelly propagated along the
timelike geodesics {\sl up to direction in spin space}. The amplitude
of the spinor $\chi^i $ is governed by the expansion
$u^\mu _{;\mu} $ of the hypersurface timelike congruence
whose tangent vector is given by
\ben
u^\mu = {1 \over m_i} g^{\mu \nu } \partial_\nu S \,.
\een
    
We also have from (\ref{Dirac}) that $\chi^i$ is
an eigenspinor of $u^\mu \gamma_\mu$.
As in flat space it follows that the spin tensor 
\ben
S_{\mu \nu} =
 {
{\bar {\chi }^i } \gamma _{[\mu } \gamma _{  \nu]} \chi ^i  \over 
{\bar {\chi } ^i   \chi ^i}       }
\een  
satisfies 
\ben
S_{\mu \nu} u^\nu =0\,.
\een  

Since $\chi ^i$ is parallelly propagated along $u^\mu$ in direction and 
since $S_{\mu \nu}$ depends only on the direction of $\chi^i$, it follows that  
the spin tensor $S_{\mu \nu} $ is parallelly transported
along the timelike congruence, just like any other perfect gyroscope.   
The geodesics are independent of the 
mass eigenvalue  $m_i $ and the polarization state given by $\chi^ i$. 
Indeed if the fermion starts off in a given polarization state 
(with the associated mass), it remains in it. 
In other words, at the L-G-J-W-K-B level, the 
weak equivalence principle, in the form of the universality of free fall,
i.e. the statement that all particles
fall in the same way in a gravitational field
independently of their mass, polarization, charge, etc., continues to  hold.    
Thus there should be no
unusual behavior in the vicinity of a spinning black hole,
or indeed in the neighborhood of any spinning system
due to the Lense-Thirring effect as suggested in \cite{Singh1,Singh2},
denied in \cite{Pal} and maintained in \cite{Singh3}.

\subsection{Neutrino mixing matrix}

\label{neutrinomix}

   Here, we briefly review the theory of the neutrino mixing matrix,
assuming that the neutrinos are Majorana.

   The lepton mixing matrix 
\cite{King} belongs to the coset $U(1)^2\backslash SU(3)$, since only
phasing of the lepton charge eigenstates $(\nu_e,\nu_\mu,\nu_\tau)$, 
but not the neutrino mass eigenstates $(\nu_1,\nu_2,\nu_3)$
(which are assumed Majorana) is possible.  One has
\ben
\begin{pmatrix} \nu_ e  \cr \nu_ \mu \cr \nu_\tau \end{pmatrix} = 
M   \begin{pmatrix} \nu _1  \cr \nu_ 2 \cr \nu_3 \end{pmatrix}\,.
\een
Thus $(\nu_e,e)$, $(\nu_\mu,\mu)$, and $(\nu_\tau,\tau)$ are doublets
under weak isospin. 

    One conventionally fixes the phases so that $M$ takes the form
\ben
M= \begin{pmatrix}
 1&0&0 \cr 0 & c_{23} & s_{23} \cr 0 & -s_{23} & c_{23}\end{pmatrix}  
\begin{pmatrix} c _{13} &0& s_{13} e^{-i \delta} \cr 0 & 1  & 0 \cr -s_{13}
  e^{i\delta}
& 0 & c_{13}\end{pmatrix}  
\begin{pmatrix} c_{12} &s_{12} &0\cr -s_{12}  & c_{12} & 0 \cr 0 & 
0 & 1\end{pmatrix}
\begin{pmatrix} e^{i\alpha _1 /2 } &0&0\cr 0 & e^{i \alpha _2 /2 }   & 0  \cr 
0 & 0 & 1\end{pmatrix} \,,
\een
where the three angles $\theta_{12}$, $\theta_{13}$, and $\theta_{23}$ 
lie in the first quadrant.

    The Jarlskog invariant for the neutrino mixing matrix, 
defined as in (\ref{Jdef}) but with $V$ 
now replaced by $M$, is again given by (\ref{Jexp1}).  Note, in 
particular, that it is independent of the phases $\alpha_1$ and $\alpha_2$.

    Experimentally, parameters of the neutrino mixing matrix are not completely known.
According to \cite{particledb}, 
\be
\sin^2 2\theta_{12} = 0.86_{-0.04}^{+0.03}\,,\qquad
0.92 < \sin^2 2\theta_{23} \le 1\,,\qquad
\sin^2 2\theta_{13} < 0.19\,,
\ee
and there is no experimental information about the Dirac angle $\delta$.
Thus, we can certainly deduce that there is an upper bound on the
Jarlskog invariant for the neutrino mixing matrix, given by
\be
|J|< 0.049\,. \label{Jneutbound}
\ee

For six different neutrino mass eigenstates, as in the seesaw mechanism, a general mixing matrix would be an element of $U(1)^5 \backslash SU(6)$, since one would diagonalize a $6\times 6$ Hermitian matrix.

\subsection{Statistics of $J$}
\label{neutrinosec}

   We have seen that the parameter space for
the neutrino mixing matrix is the six-dimensional single quotient
$U(1)^2\backslash SU(3)$, and that the Jarlskog invariant for the neutrino mixing
matrix takes the same form (\ref{Jdef}), or (\ref{Jdef2}), as it does
for the Kobayashi-Maskawa 
matrix. Therefore, all results obtained in Secs. \ref{flagman} and \ref{probdist} apply equally to the case of neutrinos.

  For completeness, we quote the results obtained in Sec. \ref{flagman}:
\ben
\langle J^2 \rangle = \fft1{720} \approx 1.389 \times 10^{-3}\,,\qquad
\langle J^4 \rangle =  \fft1{201600} \approx 4.960 \times 10^{-5}\,, \qquad 
\Delta J = \fft1{12\sqrt 5} \approx 0.0373\,.
\ee
This can be compared with the experimental bound given in (\ref{Jneutbound}).

One could also repeat the calculations of Sec. \ref{gaussian}, assuming particular values for the neutrino masses. A strong hierarchy in the neutrino masses would then presumably again lead to ``naturally'' small CP violation from the corresponding mixing matrix. Alternatively, an experimental observation of small CP violation for neutrinos would perhaps be an indication of a mass hierarchy in neutrinos. At present, neither the magnitude of CP violation nor any values of neutrino masses have been measured sufficiently accurately to allow predictions.

\section{Conclusions and Outlook}

In this paper, we analyzed the problem of finding a natural measure on a space of coupling constants, which in our case was the space of Kobayashi-Maskawa matrices, the double quotient $U(1)^2 \backslash SU(3) / U(1)^2$. We saw that the measure on this double quotient is nonunique, and we analyzed several possible choices of measure on the double quotient. One class of measures was given by squashed Kaluza-Klein measures, induced by a Kaluza-Klein reduction of a left-invariant metric on the flag manifold. Alternatively, one could take the unique measure on $SU(3) / U(1)^2$ and simply integrate over the left angles. The measure used by Ozsv\'ath and Sch\"ucking seemed not to be very well motivated from a geometric perspective.

When calculating expectation values for $J$, we found that all of the measures we considered led to rather similar statistics of $J$. In each case, the observed value was about three orders of magnitude below what one would normally expect; the observed value appears to be finely tuned. The same applied to the Ozsv\'ath-Sch\"ucking measure, an extremely squashed Kaluza-Klein measure, or a flat measure, which is just the simplest choice and not justified geometrically.

In Sec. \ref{gaussian}, we adopted the different viewpoint that the Kobayashi-Maskawa matrix should not be viewed as separate from the quark masses, but that it is really the mass matrices which are ``chosen" by a yet unknown physical mechanism. We took the observed values for the quark masses as an input and chose the simplest distribution which was able to reproduce these observed values, while inducing a different measure on the space of Kobayashi-Maskawa matrices. Assuming such a distribution, we found that the observed value of $J$ now appears very natural and not finely tuned at all. In this statistical approach, regarding the Yukawa couplings determining the mass matrices as randomly chosen seems more appropriate than separating quark masses and mixing angles. (On submittal of this article to the archive we were informed
of an earlier work~\cite{ddr}, similar in spirit to ours but using different assumptions and methods, which reaches broadly similar conclusions).

Our analysis also applies to the case of massive neutrinos, where the predictions will conceivably be tested by future experiments. In the standard theory, the Maki-Nakagawa-Sakata matrix \cite{mns} which appears is naturally an element of the single quotient $U(1)^2\backslash SU(3)$. Since the right phases do not play any role in neutrino oscillations and the relevant $J$ is independent of these phases, the calculations are identical to the ones presented here, although with the appropriate values of the $\mu$ parameters appearing in $A$ and $A'$.

In the seesaw mechanism one adds very heavy right-handed neutrinos, and the most general mixing matrix would be an element of $U(1)^5 \backslash SU(6)$. This is naturally a K\"ahler manifold, and the measure induced by the K\"ahler metric can be obtained from the analysis in \cite{Picken}. We leave a detailed treatment of this case, following our approach here, to future work.

Finally, one could analyze the effects of a fourth generation of quarks on CP violation by repeating the calculations for $4\times 4$ Hermitian matrices. If this generalization spoils the agreement with the observed $J$, one might obtain interesting lower bounds on the masses of a hypothetical extra generation of quarks.

\section*{Acknowledgemets}
GWG  would like to thank Thibault Damour, Stanley Deser, Marc Heneaux,
and John Taylor for helpful discussions and suggestions at an early stage of
part of this work.  SG acknowledges funding from EPSRC and Trinity College, Cambridge.
We thank Ben Allanach for helpful conversations and Malcolm Perry for suggesting the possible effect of a fourth generation. This research was supported in part by Perimeter Institute for Theoretical Physics.

\end{document}